%% file: mllens_revision.tex
\documentclass[reprint,prd, superscriptaddress, tightenlines, longbibliography, nofootinbib, eqsecnum, amsfonts, amsmath, floatfix, notitlepage]{revtex4-1}
\pdfoutput=1

\usepackage[utf8]{inputenc}
\usepackage{mathrsfs}
\usepackage{euscript}
\usepackage{epsfig}
\usepackage{graphics}
\usepackage{graphicx}
\usepackage{amsmath}
\usepackage{amssymb}
\usepackage{gensymb}
\usepackage{bm}
\usepackage[usenames,dvipsnames,svgnames,table]{xcolor}
\usepackage{xspace}
\usepackage{wasysym}
\usepackage{times}
\usepackage{appendix}
\usepackage{comment}
\usepackage{lipsum}
\usepackage[nolist,nohyperlinks]{acronym}
\usepackage{float}
\usepackage{simplewick}
\usepackage{natbib, ifthen}
\usepackage{hyperref}
\usepackage{longtable}

\input macros.tex

\newcommand{\beq}{\begin{equation}}
\newcommand{\enq}{\end{equation}}
\newcommand{\beqa}{\begin{eqnarray}}
\newcommand{\enqa}{\end{eqnarray}}
\newcommand{\beit}{\begin{itemize}}
\newcommand{\enit}{\end{itemize}}
\newcommand{\bem}{\begin{pmatrix}}
\renewcommand{\enm}{\end{pmatrix}}

\newcommand{\vecL}{\mathbf L}

\newcommand{\vecx}{\mathbf{x} }
\newcommand{\vecy}{\mathbf{y} }

\newcommand{\vecr}{\mathbf{r}}

\renewcommand{\Tr}{\mathrm{Tr}}

\newcommand{\lat}{\left\langle}
\newcommand{\rat}{\right\rangle}
\newcommand{\av}[1]{\lat #1 \rat}

\newcommand{\lb}{\left [}
\newcommand{\rb}{\right ]}
\newcommand{\lp}{\left (}
\newcommand{\rp}{\right )}

\renewcommand{\bem}{\begin{bmatrix}}
\renewcommand{\enm}{\end{bmatrix}}

\newcommand{\Cov}{\textrm{Cov}}

\newcommand{\D}{ {\boldsymbol \Lambda}}
\newcommand{\B}{ { B}}

\newcommand{\vecell}{ {\boldsymbol  \ell}}

\newcommand{\deflect}{{\boldsymbol{\alpha}}}
\newcommand{\deflecti}{{\boldsymbol{\alpha}^{-1}}}
\newcommand{\normdeflect}{\alpha}

\newcommand{\Dop}{D} 
\newcommand{\Dopt}{D^\dagger}
\newcommand{\St}{X} 
\newcommand{\Stdat}{X^{\rm dat}} 
\newcommand{\Stdatdag}{X^{\rm{dat},\dagger}} 

\newcommand{\X}{X} 
\newcommand{\bg}{\boldsymbol{g}}

\newcommand{\isdraft}[1]{}
\newcommand{\shrunk}[2]{#1}
\newcommand{\revision}[1]{#1}

\renewcommand{\D}{D} 
\newcommand{\lensit}{\textsc{LensIt}\xspace}
\renewcommand{\vx}{\vecx}

\newcommand{\n}{\boldsymbol{n}}
\newcommand{\hn}{\boldsymbol{n}}

\begin{document}
\newcommand{\Sussex}{Department of Physics \& Astronomy, University of Sussex, Brighton BN1 9QH, UK}
\author{Julien Carron}
\affiliation{\Sussex}
\author{Antony Lewis}
\affiliation{\Sussex}


\title{Maximum a posteriori CMB lensing reconstruction}





\date{\today}
\begin{abstract}

Gravitational lensing of the CMB is a valuable cosmological signal that correlates to tracers of large-scale structure
and acts as a important source of confusion for primordial $B$-mode polarization. State-of-the-art lensing reconstruction analyses
use quadratic estimators, which are easily applicable to data. However, these estimators are known to be suboptimal, in particular for polarization, and large improvements are expected to be possible for high signal-to-noise polarization experiments.  We develop a method and numerical code, \lensit, that is able to find efficiently the most probable lensing map, introducing no significant approximations to the lensed CMB likelihood, and applicable to beamed and masked data with inhomogeneous noise. It works by iteratively reconstructing the primordial unlensed CMB using a deflection estimate and its inverse, and removing residual lensing from these maps with quadratic estimator techniques.
Roughly linear computational cost is maintained due to fast convergence of iterative searches, combined with the local nature of lensing.
The method achieves the maximal improvement in signal to noise  expected from analytical considerations on the unmasked parts of the sky. Delensing with this optimal map leads to forecast tensor-to-scalar ratio parameter errors improved by a factor $\simeq 2 $ compared to the quadratic estimator in a CMB stage IV configuration.


\end{abstract}


\maketitle



\section{Introduction}
\indent
The large-scale structure of the Universe deflects Cosmic Microwave Background (CMB) photons by a few arcminutes, introducing a characteristic signature in the fluctuations in the CMB temperature and polarization \cite{LewisChallinor2006}.
The statistical homogeneity and isotropy of the CMB gets distorted locally, and sizeable higher-order statistics are produced.
Lensing estimators use these higher-order statistics to construct an integrated measure of the
linear mass fluctuations in the Universe that cross-correlates to all traditional large-scale structure tracers. 
After the first direct detection of lensing in the CMB by the ACT team \cite{Das:2011ak}, the SPT ~\cite{vanEngelen:2012va}, POLARBEAR \cite{Ade:2013gez}, SPTpol \cite{Story:2014hni}, BICEP2-KECK \cite{Array:2016afx} and Planck \cite{Ade:2013tyw} collaborations have also reported the detection of the lensing signal and published band-powers. The most decisive detection yet was by the Planck satellite \cite{Planck2015CMBLensing}: its full-sky coverage comes with a statistical power that simply cannot be matched at the present time.
\newline
\indent
Current measurements all use quadratic estimator techniques, first devised in optimized form by Refs.~\cite{OkamotoHu2002,OkamotoHu2003}.
The quadratic estimator uses optimally-weighted two-point statistics of the data maps to reconstruct the deflection field.
At current noise levels, this estimator is nearly optimal. The science returns from the use of more sophisticated techniques are expected to be small,
and no other type of estimator has been applied to data so far.
However, the situation will have changed by the time of CMB stage IV (CMB-S4) \cite{Abazajian:2016yjj}, if not before.
At this point the polarization instrumental noise is expected to become smaller than the $\sim 5 \mu K$ arcmin lensing $B$ mode. Barring welcome detections, lensing will become the most relevant cosmological source of confusion in the search for primordial $B$ modes \cite{Seljak:2003pn}, 
and more optimal delensing methods will become critical.

If the noise and primordial polarization $B$ mode is negligible, a well-known variable-counting argument~\cite{Hirata:2003ka} suggests that as long as the lensing is fully described by a gradient deflection remapping of the unlensed fields,
the observed lensed $E$ and $B$ fields should contain enough information to reconstruct essentially perfectly both the lensing potential and the unlensed $E$ field.
 Fundamental limits are well below near-future sensitivities, including corrections to the remapping approximation from emission angle, time delay and polarization rotation~\cite{Hu:2001yq,Lewis:2017ans}, lensing curl modes from second order post-Born lens-lens couplings~\cite{Hirata:2003ka,Pratten:2016dsm}, intrinsic non-linearities of the CMB at recombination~\cite{Mollerach:2003nq,Fidler:2014oda}, and second-order sourced vector and tensor modes~\cite{Mollerach:2003nq,Baumann:2007zm,Sarkar:2008ii,Adamek:2015mna}.
 With the last science release from the \planck\ team in sight, it therefore seems timely to revisit alternative, more optimal CMB lensing estimation. This paper presents and discusses a new implementation of a maximum a posteriori estimate of the lensing potential from CMB data.

Motivation for this work is not limited to primordial $B$ modes. The CMB lensing kernel peaks at $z \sim 2$ and overlaps the galaxy and weak lensing surveys targeting the dark sector of the Universe. The correlated information is expected to contribute to breaking important degeneracies and to help with systematics, so optimal CMB lensing mass maps will also be useful for use with large-scale structure observations. Iterative estimates may also be useful even at higher noise levels, in particular in the presence of sky cuts or wildly inhomogeneous noise maps where the analytic response of the quadratic estimator is inaccurate. An optimal estimate of the potential map might prove better than the current simpler practice of sweeping these deviations into Monte-Carlo (MC) corrections to the spectrum estimate. Finally, also looking a bit ahead, successful exploration of the lensed CMB likelihood may prove useful more generally, opening a path towards optimal joint estimation of the primary CMB and lensing potential.

It is clear how iterative estimates should work in practice, at least intuitively: delens the data using the quadratic deflection estimate, then again apply a quadratic estimator on the resulting maps, with possibly modified weights, and iterate until convergence~\cite{Smith:2010gu}. Of course, a world of potential complications lurks in the details, and no canonically best approach is known at present.
Formally, the code we present finds a maximum of the posterior probability density function (PDF) of the lensing potential. As such, it is similar in spirit to the first iterative estimator proposed for temperature reconstruction \cite{Hirata:2002jy} and polarization \cite{Hirata:2003ka}. A maximum likelihood approach to lensing reconstruction is also discussed, without implementation, in the review article Ref.~\cite{Hanson:2009kr}. In contrast to Refs.~\cite{Hirata:2002jy,Hirata:2003ka}, our implementation can be considered exact, in the sense that it maximizes the relevant functions without introducing approximations, under the assumption of Gaussian unlensed CMB, noise and deflection fields. It can also account for beams, sky cuts and other non-ideal effects.

The quadratic estimator has the convenient property of being relatively straightforward. It can be implemented using a small number of harmonic transforms \cite{DvorkinSmith2009, Planck2015CMBLensing}, keeping the overall numerical cost under control (dominated in the \planck\ analysis by the cost of the inverse-variance filtering step).  It seems unavoidable that alternative more-optimal approaches must be substantially more costly, and our implementation is no exception. At each iteration step, maximum a posteriori unlensed CMB maps are produced, under the assumption that the current deflection estimate is the correct one. This operation in effect solves a large $N_{\rm{pix}} \times N_{\rm{pix}}$ set of linear equations, and must be itself performed via an iterative method, each step involving a fair number of lensing operations, even in the absence of sky cuts or other non-ideal effects.

Nevertheless, lensing and lensing reconstruction have the advantage of being very local in position space. All operations scale linearly with the number of resolution elements, or follow the cost of an harmonic transform, and the good convergence properties of the iterative searches proposed here keep the total computational burden under control. We also provide GPU implementations of the most expensive steps.

 We use the flat sky approximation throughout the main text. Appendix \ref{appendix:curved} describes the implementation on the curved sky, using the machinery of spin-weight spherical harmonics. The implementation is otherwise identical in all respects, though we have so far only thoroughly tested everything on the flat sky where the numerical implementation is faster. We expect the same convergence properties of iterative estimator on the curved sky: empirically, the only effect we observe increasing the area is to rescale the total execution time, which is reasonable given that lensing distortions are very much localized. Furthermore, iterative delensing will probably initially be most useful on deep observations of a small patch of sky where the flat sky approximation is accurate \citep{Abazajian:2016yjj}.

\revision{Sections~\ref{sec:descr} and \ref{sec:impl} describe the algorithm and details of its implementation respectively. Section~\ref{Sec:results} provides tests and applications. We summarize and conclude in Section~\ref{sec:summary}. }

\section{Description}\label{sec:descr}

Let us first establish some notation. Let $\vecx,\vecy$ be points on a patch of the flat sky of area $V$, and $\vecr = \vecx - \vecy$ be the separation vector. The primary, unlensed Stokes CMB fields $T,E$ or $B$ are written as $\St(\vecx)$, and $\Stdat$ denotes the observed Stokes data $T, Q$ and $U$ on the data pixels, inclusive of noise and transfer function.
	We use $a,b$ in $(0,1)$ to denote the two cartesian axes of the flat sky, and use the symmetric Fourier convention, which is closest to the traditional curved sky normalization. We use the notation $\vecell$ for multipoles of the CMB maps and $\vecL$ for the lensing maps.

We denote the primordial, unlensed CMB modes $\{T,E,B \}$ as a column matrix $\X$, with primordial spectral matrix $C^{\rm unl}_\ell$
\beq
\av{\X_\vecell \X_{\vecell'}^\dagger}= \delta_{\vecell\vecell'} C_\ell^{\rm unl}.
\enq
This matrix is diagonal with respect to multipole index, but not necessarily across $T,E,B$ indices.
Also let $\Dop$ be the deflection operation that maps these unlensed CMB modes to the real space, lensed, Stokes parameters. For instance, in temperature we may write explicitly on the flat sky
\beq
\Dop^{TT}_\vecell(\vecx) \equiv \frac 1 {\sqrt{V}} e^{i\vecell \cdot \left(\vecx + \deflect\left(\vecx\right)\right)},\quad D_\vecell^{TE}(\vecx) = D_\vecell^{TB}(\vecx) = 0.
\enq
The polarization components are similar but involves the spin-2 flat-sky harmonics.
Here, and throughout, we the use approximation that the lensed fields are entirely defined by a remapping of the unlensed fields, where $\deflect(\vecx)$ is the lensing deflection angle that relates the observed lensed field at $\vx$ to the unlensed fields at $\vecx+\deflect(\vecx)$.


\subsection{Model}
\indent The model for the CMB data $\Stdat$ on the observed pixels is given by a linear response matrix $B$ operating on  the lensed sky (which includes, for example, the effect of the instrumental beam and pixel window function), plus independent noise $n$, so that
\beq
\St^{\rm dat} = B \Dop \X + n.
\enq
The pixel-pixel covariance can be written in compact notation using a series of linear operators as follows:
\beq
\label{anisoCov}
\textrm{Cov}_\deflect \equiv \la \Stdat \Stdatdag\ra =  B \Dop C^{\rm unl} \Dopt B^\dagger + N,
\enq
where $N$ is the noise covariance matrix, which we assume is diagonal in pixel space.
Unlensed CMB fields and the noise on each pixel are assumed to obey Gaussian statistics, so the likelihood is also Gaussian. The log-likelihood is then
\beq
\label{lik}
\begin{split}
\ln p(\Stdat | \deflect) &= -\frac 12 \Stdat\cdot \Cov_\deflect^{-1}\Stdat - \frac 12 \det \Cov_\deflect.
\end{split}
\enq
We need to use a prior on the statistics of the deflection field to regularize the large number of poorly-constrained small-scale modes. The $\Lambda$CDM CMB lensing potential $\phi$ is expected to be nearly linear, so choosing Gaussian field statistics for $\phi$ is a natural choice, and will likely remain accurate in the foreseeable future on the scales where the lensing potential can accurately be reconstructed. Using a Gaussian prior on the signal does not prevent reconstruction of any non-Gaussian signal that may actually be present (as expected from non-linear structure growth and post-Born lensing~\cite{Namikawa:2016jff,Pratten:2016dsm}).

We assume pure gradient lensing deflections, in which case the log-posterior becomes, up to irrelevant constants,
\beq
\ln p (\phi | \Stdat) = \ln p(\Stdat | \phi) - \frac 12 \sum_\vecL \frac{\phi_\vecL^2}{C_\vecL^{\phi\phi}},
\enq
where the likelihood is given by Eq.~\ref{lik} with $\deflect = \nabla \phi$.
Curl and joint curl / gradient reconstruction is very analogous, but should be of limited physical relevance in the foreseeable future, mainly serving as a consistency check on the gradient reconstruction analysis.
An interesting first prospect would be the detection of the post-Born curl signal, forecast to be marginally detectable in the bispectrum with CMB-S4~\citep{Pratten:2016dsm}, to which this methodology could also be applied.

\subsection{Gradients}
To maximize the log-posterior we consider the derivative of the log-posterior with respect to the deflection.
The total gradient, $g$, splits naturally into three pieces:
\beq
g_a^{\rm tot} \equiv \frac{\delta \ln p(\deflect|\Stdat  )}{\delta \normdeflect_a(\n)}
= g^{\rm QD}_a - g^{\rm MF}_a + g^{\rm PR}_a,
\enq
 one from the quadratic part of the likelihood ($\bg^{\rm QD}$), one from the likelihood covariance determinant ($\bg^{\rm MF}$, the mean-field), and one ($\bg^{\rm PR}$)  from the prior. The choice of the odd sign of $\bg^{\rm MF}$ is more natural and becomes clear later on.
The prior gradient is straightforward to evaluate assuming Gaussian statistics.

The gradients of the likelihood are first calculated in real space, with the gradients with respect to the two cartesian components of the deflection giving
\beq
\frac{\delta \ln p(\Stdat | \deflect )}{\delta \normdeflect_a(\n)}= g^{\textrm{QD}}_a(\n) - g^{\rm MF}_a(\n).
\enq
These are then rotated to harmonic space to give the gradient and curl components. The piece quadratic in the data
\beq
g^{\textrm{QD}}_a(\n)  = \lb V_\deflect \Stdat\rb^{i} (\n) \lb W^{a}_{\deflect} \:\Stdat\rb_{i} (\n),
\label{gQD}
\enq
is made up of two legs with data weights
\beq
V_\deflect = B^\dagger \Cov^{-1}_\deflect,\quad W^a_{\deflect} = D \:\nabla_a C^{\rm unl}\:D^\dagger \:\B^\dagger \Cov_\deflect^{-1}.
\label{optweights}
\enq
The gradient matrix $\nabla_a C^{\rm unl}$ is block diagonal in harmonic space with blocks $i\vecell_a C^{\rm unl}_\vecell$.
These weights are identical, in the absence of deflection, to the (unnormalized) traditional Minimum Variance (MV) lensing quadratic estimators evaluated with unlensed spectra.

Both legs  of the quadratic estimator (the two terms in Eq.~\eqref{gQD}) can be written in terms of reconstructed unlensed CMB modes, as follows. Consider the most probable primordial CMB modes $\X^{\rm WF}_{\deflect}$ given the data, under the assumption that $\deflect$ is the true deflection field, and that they are Gaussian fields with power $C_\ell^{\rm unl}$. The maximum a posteriori (MAP) unlensed CMB maps are formally given by the Wiener-filtered data\footnote{For a signal seen under a linear response $s^{\rm dat} = R s^{\rm true} + n$ the maximum a  posteriori reconstructed signal $s$ is given assuming Gaussian statistics by $s^{\rm WF} = (S^{-1} + R^t N^{-1}R)^{-1} R^t N^{-1}s^{\rm dat} =  S R^t (R S R^t + N)^{-1}s^{\rm dat}$}
\beq
\label{MAP}
X^{\rm WF}_\deflect \equiv C^{\rm unl} \Dopt B^\dagger \Cov_\deflect^{-1}\St^{\rm dat}.
\enq
The leg $W^a_\deflect X^{\rm dat}$ of the quadratic piece is then simply the deflected gradient
of these maps
\beq
\label{Leg1}
W^{a}_{\deflect} \:\St^{\rm dat}(\vecx) = \Dop \nabla_a \X^{\rm WF}_{\deflect}(\vecx).
\enq
The other leg can be written as the inverse-noise-weighted residual between the data and how the inferred primordial modes are predicted to appear:
\begin{eqnarray}
\label{Leg2}
V_\deflect \St^{\rm dat}
&=& B^\dagger \Cov_\deflect^{-1} \St^{\rm dat}
\nonumber\\
&=& B^\dagger N^{-1}\left( \Cov_\deflect - B D C^{\rm unl} D^\dag B^\dagger\right)\Cov_\deflect^{-1} \St^{\rm dat}
\nonumber \\
&=&B^\dagger N^{-1}\lb \St^{\rm dat}  - B \Dop \X^{\rm WF}_{\deflect}\rb.
\end{eqnarray}
The calculation of these two terms is simple once the maps $\X^{\rm WF}$ are reconstructed.
Our implementation is discussed in Sec.~\ref{Sec:MAP}.

The second part of the likelihood gradient is the contribution from the mean field
\beq\label{MF}
\begin{split}
g^{\rm MF}_a(\n)  &= \frac 12 \frac{\delta \ln \det \Cov_\deflect}{\delta \normdeflect_a(\vecx)} =  \av{g_a^{\textrm{QD}}(\n)}.
\end{split}
\enq
The average here is over realizations of the data, with displacement $\deflect$ held fixed. The second equality follows from observing that the first variation of a log-likelihood always vanish in the mean, and that $g^{\rm MF}_a(\n)$ itself is independent of the data (and hence is equal to its expectation).
The mean field serves the same purpose here as for the traditional quadratic estimator: to subtract the known sources of anisotropy from the quadratic estimate. It depends on the current estimate of the deflection, because $\deflect$ at this iteration acts as a known source of anisotropy when measuring residual lensing at the next iteration. Implementations are discussed in Sec.~\ref{Sec:MF} and Appendix~\ref{appendix:MF}.


 \begin{figure*}[htp]
 \centering
\shrunk{\includegraphics[width = \textwidth]{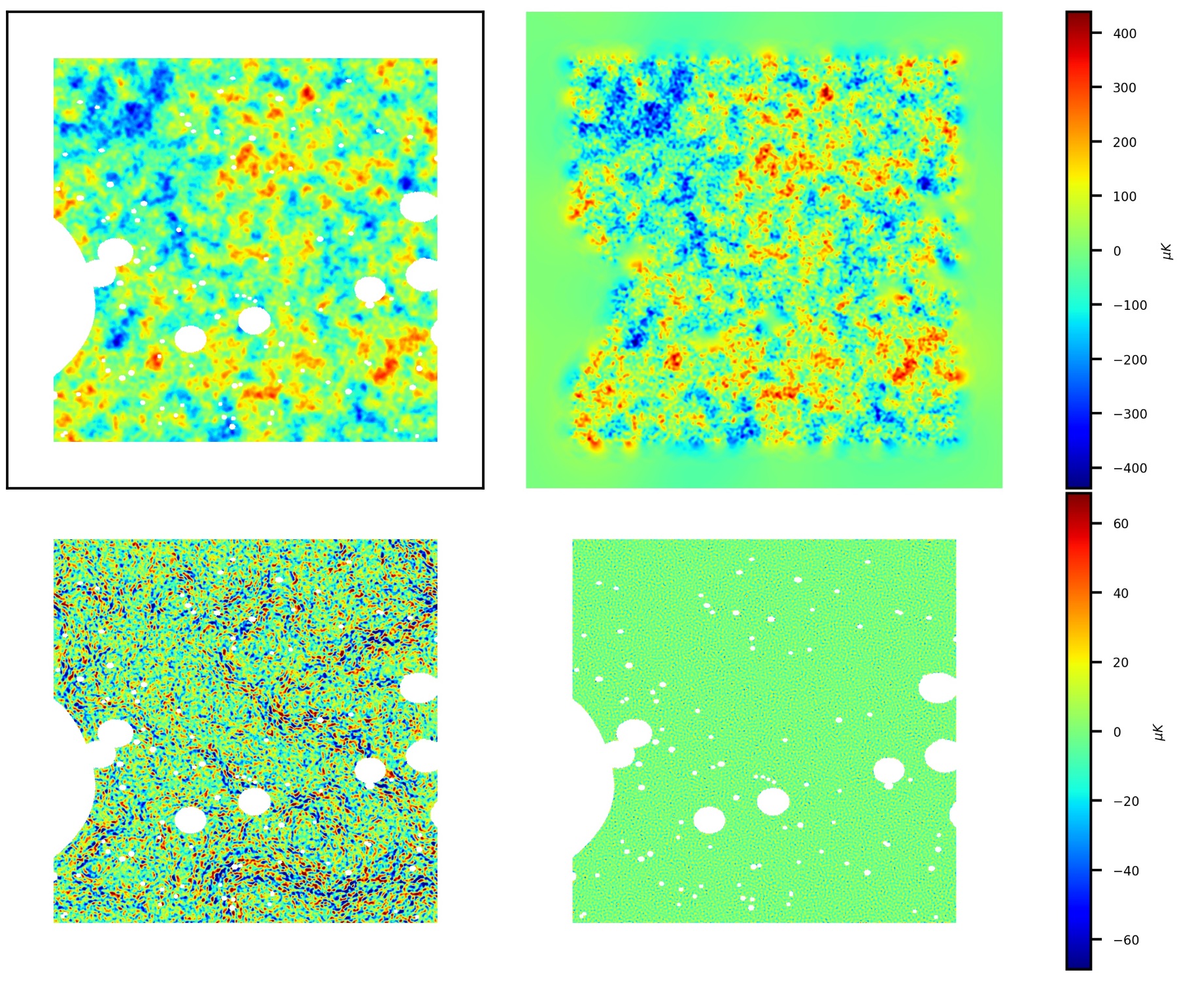}}
{\includegraphics[width = \textwidth]{figs/filt_plot_Planck_idealjet.jpeg}}

 \caption{\label{fig:filt}
 A demonstration of how our Wiener filter produces optimal (maximum a posteriori) estimates on the unlensed CMB maps from masked data. The simulated temperature map is comparable to a \planck\ configuration, and we use the exact input simulated deflection and exact input unlensed $C_\ell^{TT}$ spectrum in the filter. The upper-left panel shows the simulated masked temperature data map, with a homogeneous noise level of 35 $\mu \rm K$-arcmin and a beam FWHM of $7$-arcmin. The (unapodized) mask is built out of a portion of the public \Planck\ lensing mask, to which we have added a band surrounding the patch on all sides. The upper right panel shows the reconstructed unlensed map $T^{\rm WF}$. The residual to the true input CMB map ($T^{\rm WF} - T^{\rm input}$) is shown on the lower-right panel.
  The lower-left panel shows the residual (on the unmasked pixels) of the result obtained when the Wiener filter instead uses no deflection but the lensed CMB spectrum in place of the unlensed spectrum (as in the standard quadratic estimator). These residuals are several times larger in magnitude (the same colour scale is sometimes saturated), and display the anisotropic swirly patterns generated by the pure gradient deflection field.} \end{figure*}
\section{Implementation}\label{sec:impl}
This section presents some details of our implementation.
The main numerical difficulty lies in the calculation of the Wiener-filtered modes $X^{\rm WF}_\deflect$ in the presence of the deflection $\deflect$ (and anisotropic noise, masks, etc.). This is discussed in Sec.~\ref{Sec:MAP}. The Wiener-filtering operation  requires the inversion of the deflection field, described in Sec.~\ref{Sec:lensing}, and the mean-field evaluation is discussed in Sec.~\ref{Sec:MF}. We also make use of curvature information to improve convergence, as discussed in Sec.~\ref{Sec:curvature}. Finally, we describe our choice of starting point in Sec.~\ref{Sec:start}, and summarize the workflow of the method in Sec.~\ref{Sec:workflow}.
\subsection{Reconstruction of the unlensed CMB \label{Sec:MAP}}

The MAP estimate of the unlensed CMB given a deflection field $\deflect$  is formally given by Eq.~\ref{MAP}, and typically requires solving a large system of linear equations. This is not straightforward even in the absence of sky cuts or other non-ideal effects, since the deflection field breaks isotropy so that the harmonic transforms do not diagonalize the system. However, for all realistic situations we have investigated, we found that the additional complication of the deflection field was minor in comparison to (typically highly anisotropic) sky cuts.
\newline
\indent
Our implementation is as follows. We first transform Eq.~\ref{MAP} to the following form
\beq
\label{filter}
X^{\rm WF}_\deflect = \lb \left( C^{\rm unl} \right)^{-1} + \Dopt B^\dagger N^{-1} B \Dop \rb^{-1} \Dopt B^\dagger N^{-1}\Stdat,
\enq
and solve for the large inverse in brackets with conjugate gradient descent. Here one should understand the bracketed matrix to act on the space of non-zero unlensed CMB modes. There is no ambiguity regarding the unlensed spectra $(C^{\rm unl})^{-1}$, since all modes in $X^{\rm WF}$ are exactly zero when they correspond to fiducial $C_\ell^{\rm unl}$ that are zero: the Wiener filter builds the maximum a posteriori $\X$ maps, hence these vanish whenever the prior variance $C^{\rm unl}$ does. We found Eq.~\eqref{filter} to be more efficient than other possible ways to perform the mask deconvolution, and it also is more efficient when the lensing operations are the only source of complications.

For the noise matrix $N$ we use an input variance map that is diagonal in pixel space. The noise can be inhomogeneous, and we can also add to the noise matrix a set of templates
which are projected out. This is useful for instance to account for poorly understood low-$\ell$ noise, or to project out any templates for galactic dust.
 Using conjugate gradient descent requires a reasonably fast way to apply $ \left(C^{\rm unl} \right)^{-1} + \Dopt B^\dagger N^{-1} B \Dop $ to vectors $\X$. The first term is diagonal in harmonic space and poses no problem. The second term requires application of the lensing operators $\D$ and $\D^\dagger$, and of the inverse noise matrix. From Eq.~\ref{Dt}, applying the lensing operator to a map is simply achieved by harmonic transforms followed by lensing of the resulting map. As discussed in more detail in Sec.~\ref{Sec:lensing}, $\Dopt$  also involves the inverse harmonic transform and a delensing operation (lensing with the inverse deflection field). It therefore has the same complexity as forward lensing, provided the inverse deflection field has been precomputed. The inverse noise matrix is simple under the assumption that it is diagonal in pixel space. The inclusion of templates is only a minor complication as long as there are only a reasonable number of them.
 Finally, assuming isotropic beams, the beam operations are fast in harmonic space.

All in all,  application of the bracketed matrix in Eq.~\ref{filter} requires 4 harmonic transform and 2 lensing operations, multiplied by 1, 2 or 3 for temperature only, polarization only or joint reconstruction respectively. Lensing of maps (or the displacement inversion) is not a cheap operation, but the cost scales linearly with the number of pixels and can easily be parallelized. Our implementations, including a GPU implementation, are discussed in Sec~\ref{Sec:lensing}.

The use of a good preconditioner is mandatory for convergence in acceptable time, especially when dealing with masked maps. We use a multigrid preconditioner following Ref.~\cite{Smith:2007rg}, where a set of working resolutions is set up so that lower-resolution
inverses are used to precondition those at higher resolution.
Specifically, we extend the \textit{qcinv} package\footnote{\url{https://github.com/dhanson/qcinv}} by Duncan Hanson to include the lensing operations. At the lowest resolution stage, we use a dense preconditioner. We offer no unique recipe of a good multigrid chain as performance appears to depend substantially on the specific configuration.  The solution $X^{\rm WF}_{\deflect_N}$ obtained at iteration $\deflect_N$ can however be used as starting point for iteration $N + 1$, which significantly speeds whole process as $\deflect_N$ settles down to the converged estimate.

Finally, we note that in ideal situations where $\deflect$ is the only source of anisotropy, the use of a simple diagonal preconditioner is much faster, with no need to resort to a multigrid solution, at least up to noise levels of a CMB-S4 configuration we have been testing.

\subsection{Lensing and delensing operations \label{Sec:lensing}}
Lensing of maps is done at a resolution of 0.7 arcminutes, using a standard bicubic spline interpolation. Lensing is an expensive operation, even if easily computed in parallel, and there are a large number of maps to process until convergence is reached, and this can dominate the overall computational cost in typical runs.  We found that porting the lensing on GPU, using a GPU-optimized implementation \citep{Ruijters2008,Ruijters2012} can provide substantial speed-up. This is one of the implementations that we provide.

In addition to the forward lensing operation, the filtering step also requires applying $D^\dagger$. This is equivalent to applying the inverse deflection together with multiplication by the magnification. To see this, consider the temperature part only. From the explicit form of the operator $\D$ in Eq.~\ref{Dt} we have
\beq
\label{Dt}
\begin{split}
\lb \D^\dagger T \rb_{\vecell}&= \frac 1 {\sqrt{V}} \int d^2x \:e^{-i \vecell\cdot (\vecx + \deflect(\vecx))} T(\vecx) \\
&= \frac 1 {\sqrt{V}} \int d^2x e^{- i \vecell \cdot \vecx} |M_{\deflecti}|(\vecx) T(\vecx + \deflecti(\vecx)).
\end{split}
\enq
The second line follows from the first after the obvious change of variable $\vecx \rightarrow \vecx + \deflect(\vecx)$, where $|M|$ is the magnification matrix determinant that accounts for the change of volume element in these new coordinates:
\beq
\lb M_\deflect \rb_{ab}(\vecx) = \delta_{ab}  + \frac{\partial \normdeflect_a}{\partial x_b}(\vecx).
\enq
The inverse deflection $\deflecti$ is defined by the condition that points deflected by $\deflect$ are remapped to themselves
\beq
\vecx + \deflect(\vecx) +  \deflecti(\vecx + \deflect(\vecx)) \equiv \vecx.
\enq
Eq.~\ref{Dt} has the simple form of the harmonic transform of the delensed temperature map, multiplied by the magnification of the inverse deflection. The generalization to polarization is immediate.

We therefore need to obtain the inverse deflection field. While some approximation to the inverse deflection is possible given the noise levels of current data \citep{Anderes:2014foa,Carron:2017vfg}, we found the exact inversion is always well-behaved for a $\Lambda$CDM displacement, is always in the weak-lensing regime, and is not a bottleneck for our reconstruction. The inversion is a very localized operation which is easily parallelized, for which we use a simple real space Newton-Raphson scheme on a high-resolution grid. Specifically, following Ref.~\cite{Carron:2017vfg}, we solve iteratively for $\deflecti(\n)$ using
\beq
\begin{split}
&\deflecti_{N + 1}(\n)
= \deflecti_N(\n) \\ &- M_\deflect^{-1}(\n + \deflecti_N(\n)) \cdot (\deflecti_N(\n) + \deflect(\n + \deflecti_N(\n))).
\end{split}
\enq
In practice, we use the same $0.7$ arcmin grid spacing that we use for the lensing operations, in which case 3 iterations starting from $\deflecti = 0$ are enough for essentially exact inversion of a typical $\Lambda$CDM deflection field. Typical resulting r.m.s. fractional residuals on the deflection amplitude are as low as $2\cdot10^{-5}$. For lensing reconstruction in a realistic situation, the forward deflection is much smoother in comparison to a typical $\Lambda$CDM deflection owing to the prior effectively filtering out many small-scale modes, and coarser resolutions may also be used.
\subsection{Mean field  evaluation \label{Sec:MF}}
Provided with a large number of data simulations, the mean field may be evaluated using
\beq
\bg^{\rm MF}(\n)= \av{\bg^{\rm{QD}}(\n)},
\enq
i.e.,  by repeating the quadratic estimate on a number $N_{\rm MC}$ of independent simulations of the data maps and averaging to get the mean field. In practice, each of these naive estimates of $\bg^{\rm MF}$ has spectrum $|\bg^{\rm QD}|^2$ and a large resulting Monte-Carlo (MC) noise, containing the signal and noise parts of $\bg^{\rm QD}$.
On small scales this noise is typically much larger than  $\bg^{\rm MF}$, so for a reasonable number of MC simulations the mean field subtraction would effectively be adding noise with power $|\bg^{\rm QD}|^2/ N_{\rm MC}$ to the estimation of the gradient on these scales. It is therefore desirable to obtain better ways to estimate the mean field. We suggest two types of trick to accelerate convergence of the mean-field estimation.

The first simply subtracts some of the Monte Carlo noise by subtracting a mean field calculated using an isotropic approximation to the data likelihood, using the same random phases for the simulations. This introduces no bias, since by isotropy the correction vanishes in the mean, but has the virtue of cancelling part of the MC noise where the anisotropy is mild.  For example, the isotropic approximation could consist of recalculating the same quadratic estimate but setting $\deflect$ to zero in the weights (in the absence of other non-ideal effects), or using a simulation extended to full sky in the presence of sky cuts.

The second trick is to modify the weights of the quadratic estimator, in a way that keeps its expectation value (i.e., the mean field) constant. This is discussed in more detail in Appendix \ref{appendix:MF}. Combined, these tricks can lead to orders of magnitude decrease of the MC noise on the mean-field estimator, drastically reducing the number of simulations required for the same target accuracy.

In principle, different random phases must be used for the simulations at each iteration step: usage of the same phases at each step causes artificial convergence of the iteration towards what is an approximation to the true posterior. This approximation might still be fairly good, however, if enough simulations are used. At any given scale it is the mean field MC noise that sets the accuracy at which the true maximum a posteriori deflection solution can be determined. We refer to this later on as the MC noise floor.

\subsection{Curvature  \label{Sec:curvature}}

Finally, to perform an efficient search for the optimal point
we need the curvature of the likelihood as well as the gradient.
Specifically, to perform efficient Newton-type iteration across parameter space, the inverse curvature is needed. Curvature matrices such as
\beq \label{Curv}
\left[H^{-1}\right]^{ab}_{\vecL \vecL'} \equiv - \frac{\delta^2 \ln p(\Stdat | \deflect)}{\delta \normdeflect_\vecL^a \:\delta \normdeflect^{b,*}_{\vecL'}}
\enq
can never be evaluated exactly in reasonable time. We proceed as follows: starting with an initial isotropic guess, $H_0$, we perform a rank two update to $H$ every time we move across parameter space.  At each iteration, two maps are saved to disk and can be used to apply recursively the inverse curvature matrix to any vector. We use the limited memory Broyden-Fletcher-Goldfarb-Shanno (L-BFGS) update \citep{Nocedal1980}, for which
\beq
H_{N + 1} = (1 + \rho s y^t)H_N(1 + \rho y s^t) - \rho ss^t,
\label{BFGS}
\enq
with
\beq
s(\n) = \phi_{N + 1}(\n) - \phi_N(\n), \quad y(\n) = g^{\rm tot}_{N + 1}(\n) - g^{\rm tot}_N(\n),
\enq
and $\rho = 1 / y^t s$.
Built in this way, the inverse curvature takes into account non-Gaussian and realization-dependent aspects of the likelihood, and we found can dramatically improve the convergence properties of the iterative search. Since the inverse curvature approximates the covariance, it can also be used to assess the width of the posterior density function, giving us approximate confidence regions for free at the end of the iterative process.

\subsection{Starting point}\label{Sec:start}
If the posterior density were exactly Gaussian, a single Newton step starting from $\deflect \equiv 0$ would bring us directly to the optimal solution.
This solution matches (neglecting the difference between the realization-dependent curvature and its average) the Wiener-filtered quadratic estimator calculated with unlensed weights~\cite{Okamoto:2003zw}. However, we use lensed weights since they provide a better quadratic reconstruction~\cite{Hanson:2010rp}, especially on large scales. Explicitly, we use
\beq
\label{start}
\deflect_0(\vecL) =  \frac{C^{\phi \phi}_L}{C_L^{\phi\phi} + N^{0,\rm len}_L} i\vecL \:\hat \phi^{\rm qest}(\vecL),
\enq
where $ N^{0,\rm len}_L $ is the Gaussian reconstruction noise of the quadratic estimator, calculated with the lensed CMB spectra. $N^{0,\rm len}_L$ is calculated using its real space flat-sky representation (see Appendix~\ref{appendix:MF}).  The quadratic estimator is the minimum variance (MV) estimator built with the set of maps considered: $T$ alone, $Q$ and $U$ polarization alone, or the three Stokes maps in combination.  The filtering step is described in Sec.\ref{Sec:MAP}, with the $(\deflect = 0,C_\ell^{\rm len})$ Wiener filter using an input noise variance map and fiducial beam transfer function.
In temperature, there are known ways to optimize the weights further~\cite{Peloton:2016kbw}, but Eq.~\eqref{start} works well for our purposes.
\subsection{Summary of the workflow}\label{Sec:workflow}
We are now in position to summarize and describe the workflow of the iterative search for the maximum a posteriori point.
Initially, the displacement $\deflect_0$ is set at the Wiener-filtered quadratic estimator as described in Sec.~\ref{Sec:start}. To get the optimal reconstruction we apply the following steps recursively until satisfactory convergence is reached:
\begin{enumerate}
\item The displacement $\deflect_N$ is inverted to give $\deflect^{-1}_N$ and cached.
\item With the deflection and its inverse, the delensed maps $X^{\rm WF}_\deflect$ are obtained with the $(\deflect_N,C_\ell^{\rm unl})$ Wiener filter using multigrid-preconditioned conjugate gradient inversion. This is the most expensive step of the whole process by some margin.
\item With the delensed CMB  at hand, the quadratic part of gradients are calculated from Eqs.~\eqref{Leg1} and \eqref{Leg2}.
\item In parallel (unless neglected, or some other approximation scheme is used), the mean field contributions to the gradient are calculated by repeating steps 2-3 on a number of simulated maps, using the tricks discussed in Sec.~\ref{appendix:MF}.
\item The total gradient $\bg_N$ is then obtained, the inverse curvature updated according to the BFGS scheme of~\eqref{BFGS}, and the displacement $\deflect_{N +1}$ is found along the Newton descent direction:
\beq
\label{Newton}
\deflect_{N + 1} = \deflect_N +  \lambda \: H_N \bg_N.
\enq
The parameter $\lambda$ helps improve convergence. For CMB-S4-like configurations, we picked $\lambda = 1/2$ at all steps, whereas the full Newton step $\lambda = 1$ can safely be used at higher noise levels.
\end{enumerate}
  \begin{figure}[htp]
 \includegraphics[width = 0.5\textwidth]{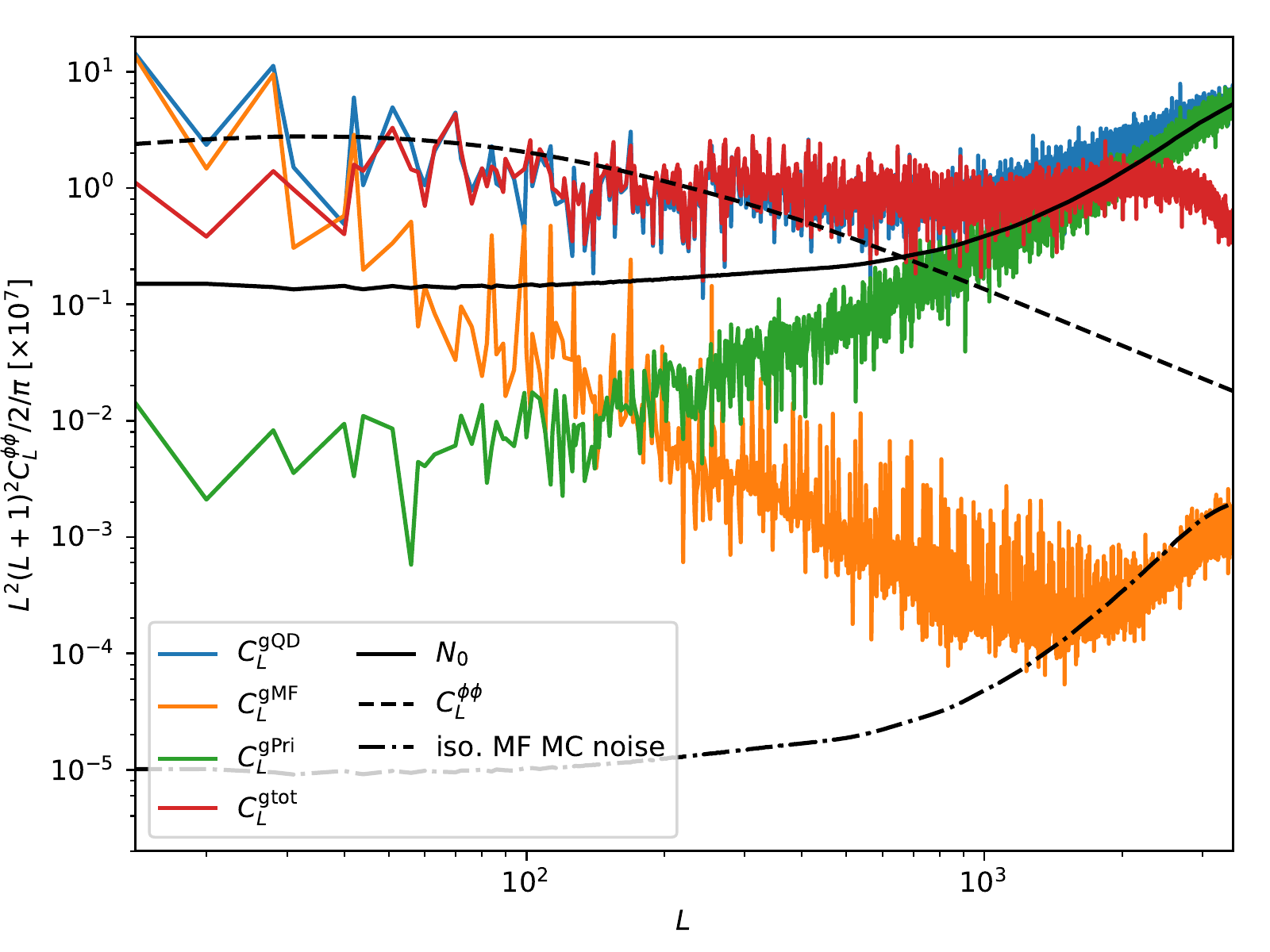}
 \includegraphics[width = 0.5\textwidth]{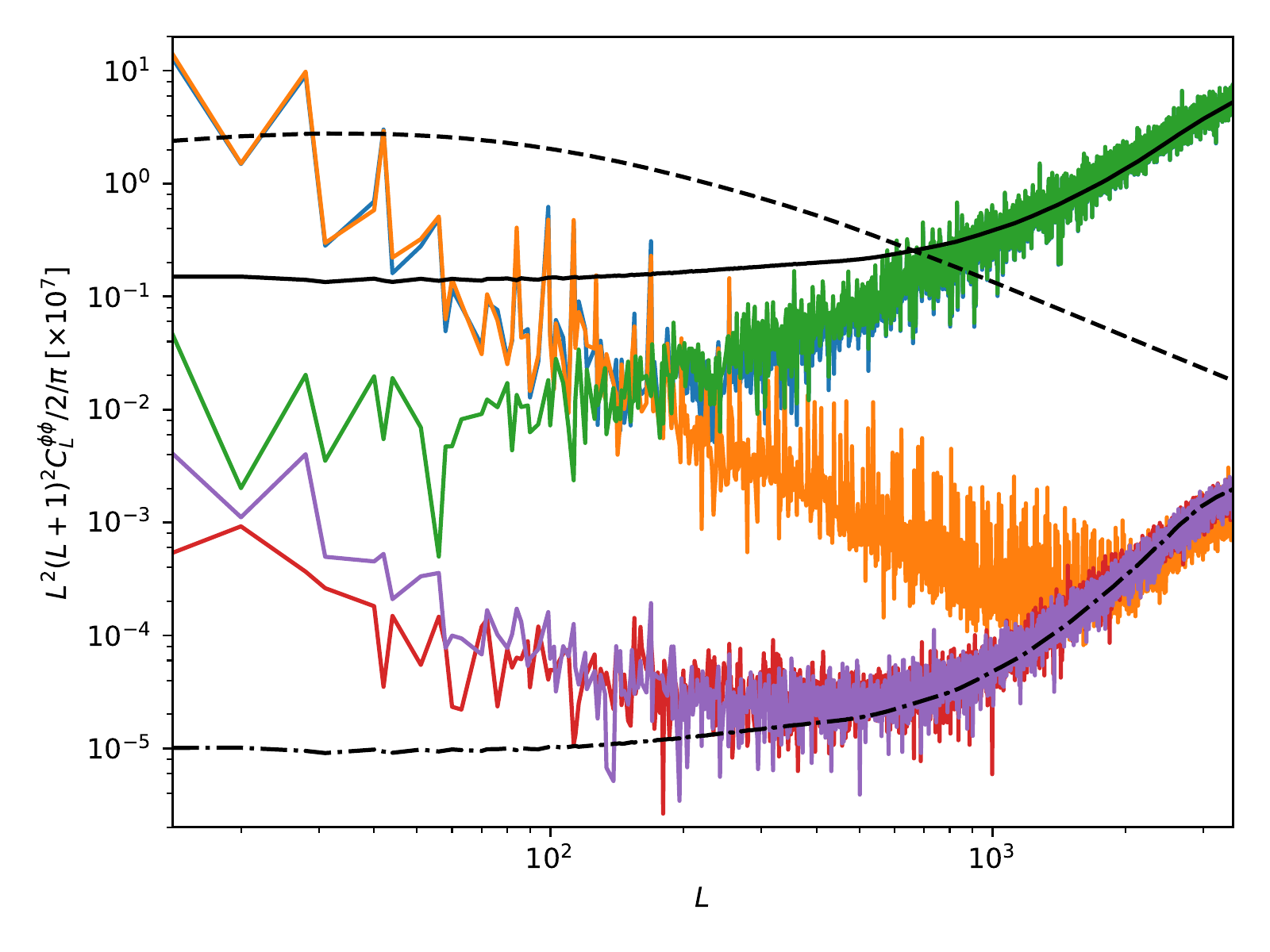}
 \caption{\label{fig:gplots}Power spectra of the three gradients $\bg^{\rm QD}$(blue), $\bg^{\rm MF}$(orange) and $\bg^{\rm PR}$(green) for lensing reconstruction from polarization on a masked patch, together with the total gradient spectrum (red). The algorithm works by reducing the red curve as much possible to find the most probable lensing map. The mask is shown in Fig.~\ref{fig:filt}, and causes the large contribution from the mean field at low multipoles. All curves are normalized to the quadratic estimator normalization $N^0_L$. The upper panel shows the gradient spectra at the first iteration step, where the deflection is the Wiener-filtered quadratic estimator, and the lower panel shows the result after 20 iterations. At this point, the gradient has hit the mean-field MC noise floor (purple on the lower panel) on all scales and the solution cannot be improved by more iterations. The MC noise floor is mean-field estimator dependent and inversely proportional to the number of simulations used (here 511 per iteration). The dot-dashed line shows predictions for the MC noise floor built from an isotropic likelihood, which are inaccurate at low multipoles because of the sky cuts. At low multipoles, the improved reconstruction relies on accurately cancelling the mean field contribution, but on intermediate scales the decrease in the quadratic estimate is immediately visible.}

 \end{figure}
 \begin{figure*}[htp]
\shrunk{
\includegraphics[width = 0.33\textwidth]{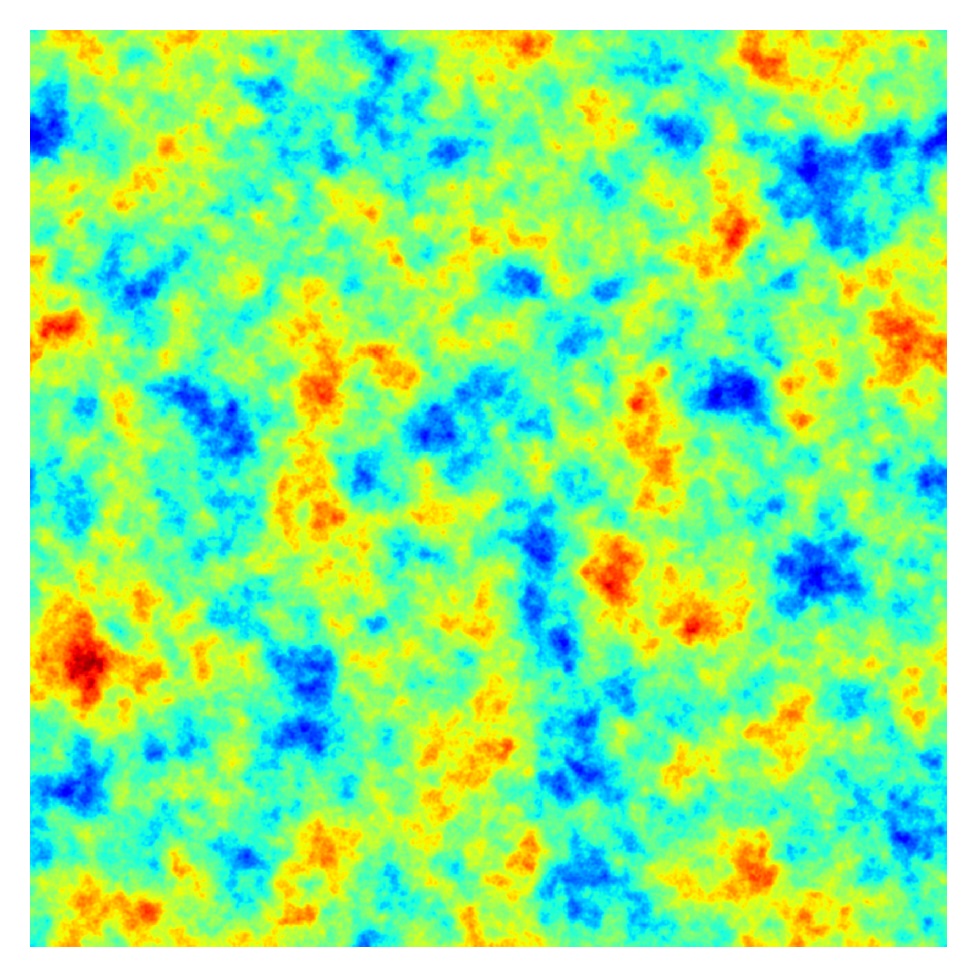}
\includegraphics[width = 0.33\textwidth]{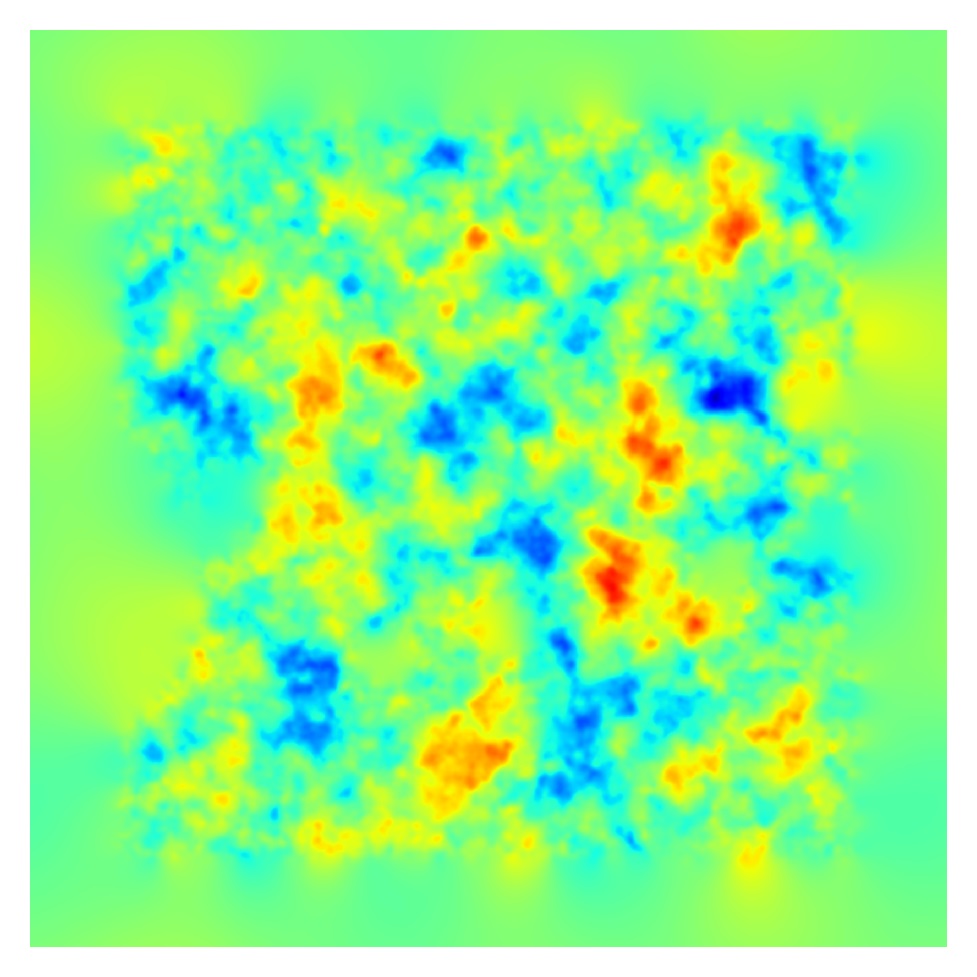}
\includegraphics[width = 0.33\textwidth]{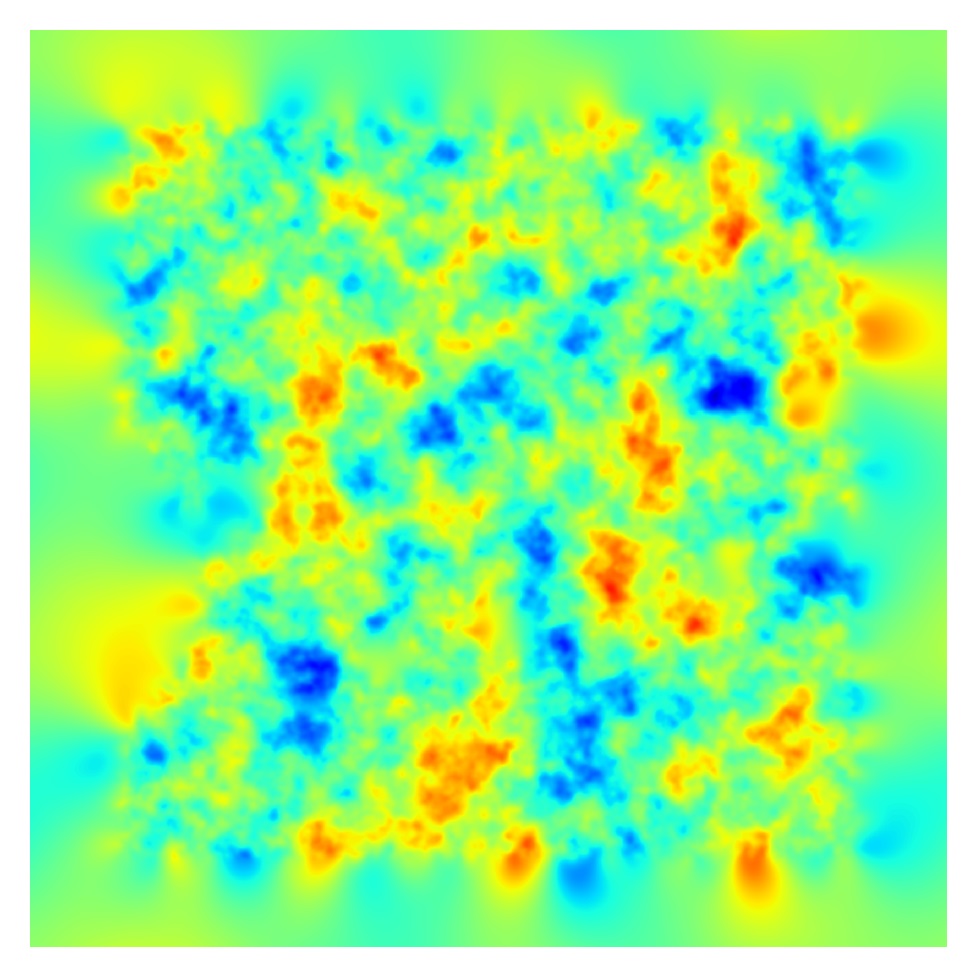}
\includegraphics[width = 0.497\textwidth]{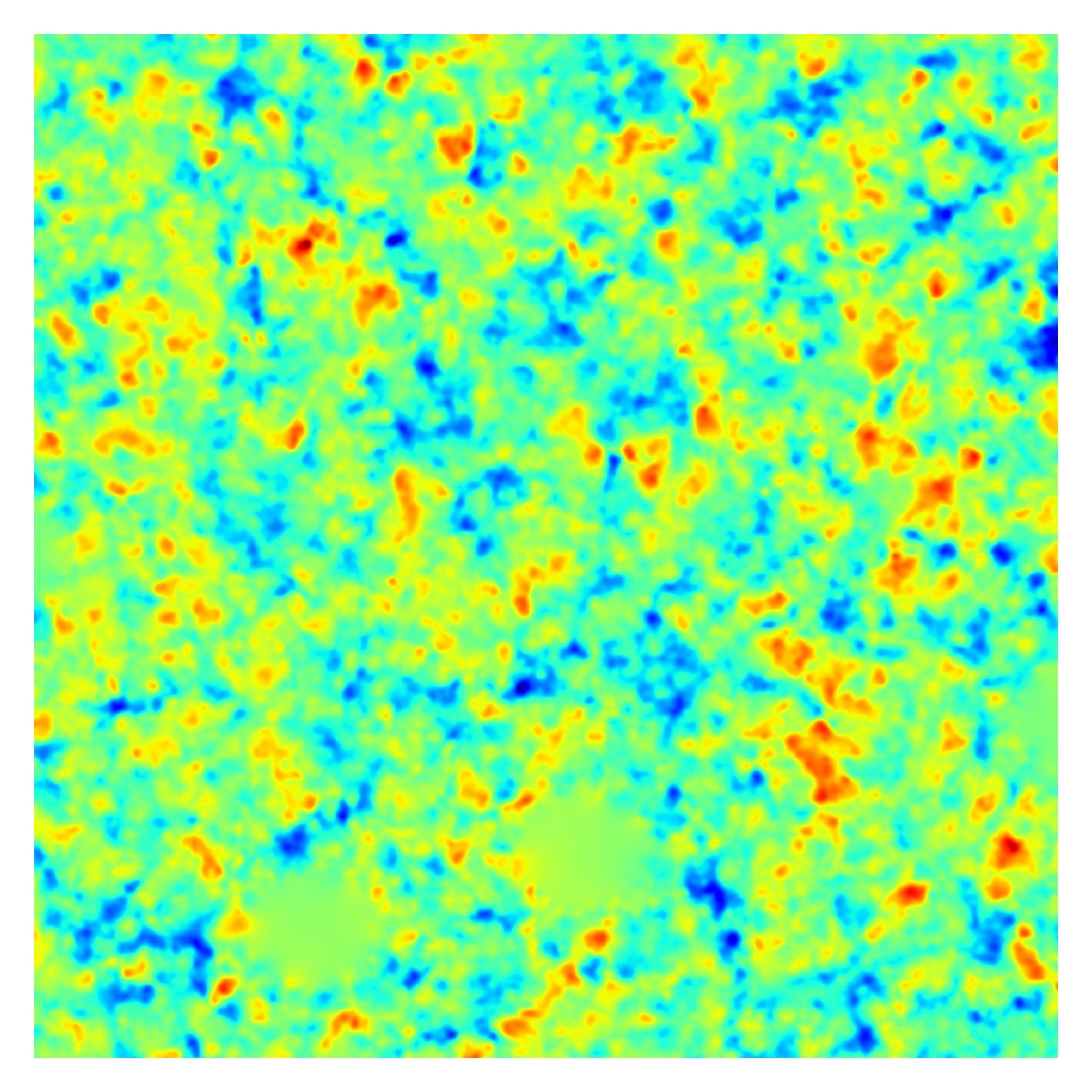}
\includegraphics[width = 0.497\textwidth]{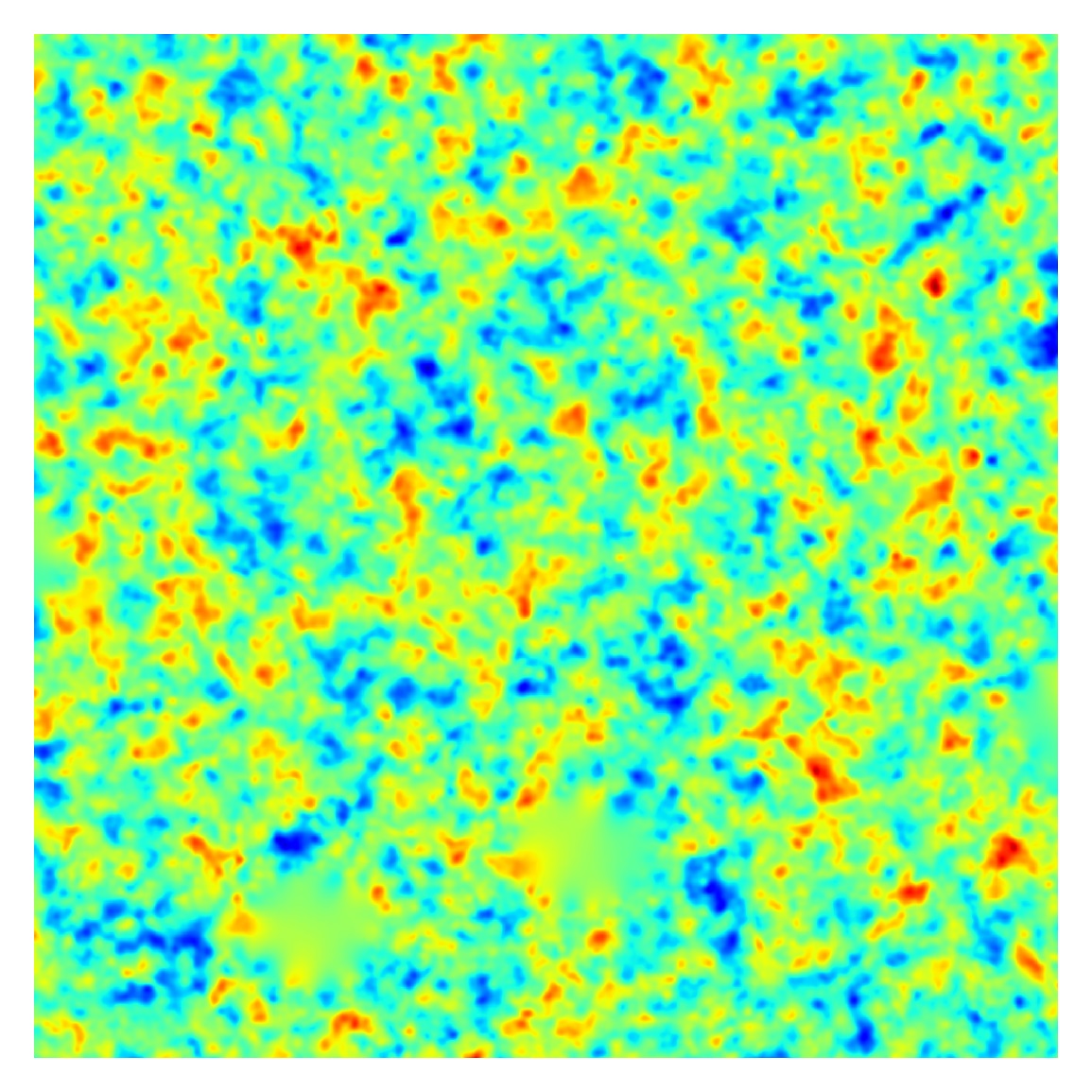}
}{
\includegraphics[width = 0.33\textwidth]{figs/QUrec_input_jetalpha.pdf}
\includegraphics[width = 0.33\textwidth]{figs/QUrec_qest_jetalpha.pdf}
\includegraphics[width = 0.33\textwidth]{figs/QUrec_itmax_jetalpha.pdf}
\includegraphics[width = 0.497\textwidth]{figs/QUrec_qest_centerWFjetkappa.pdf}
\includegraphics[width = 0.497\textwidth]{figs/QUrec_itmax_centerWFjetkappa.pdf}
}
 \caption{\label{fig:QUrec}
 The top-left panel shows a simulated lensing field used as input to a lensing reconstruction analysis. The input Stokes $Q$ and $U$ maps are masked with the same mask shown on Fig.~\ref{fig:filt}, and  we assume a polarization-sensitive experiment having polarization noise level of $1.5\cdot \sqrt{2}\mu$K-arcmin and 3 arcmin FWHM beam.  The middle panel shows on the same colour scale the Wiener-filtered quadratic estimate, which is the starting point of the iterative solution for the maximum a posteriori solution, Eq.~\ref{start}. The top-right panel shows the converged solution.
 The top three panels show the displacement-like scalar field with transform $|\vecell| \hat \phi_\vecell$, Eq.~\eqref{dx}.
 The lower two panels show the convergence maps $\kappa(\vecell) = -\frac 12 \ell^2 \hat \phi(\vecell)$ (only the central regions covering one fourth of the map) for the quadratic estimator (left) and iterative solution (right). The iterative solution can resolve structure down to smaller scales, and improvement can also be seen in the masked regions.}
 \end{figure*}
\section{Results}\label{Sec:results}

For this preliminary investigation we report three tests of our reconstruction method. First, using a simulated lensed map similar current \planck\ public data, we demonstrate use of the Wiener-filtering procedure to extract the maximum a posteriori estimate of the unlensed CMB. Second, we simulate a lensing reconstruction from polarization on a masked field, with noise levels of next-generation CMB experiments. \revision{In its simplest incarnation, the algorithm does not directly provide a delensed primordial $B$-mode map, but the iteratively reconstructed potential map can still be used to delens the observed $B$ modes. Third and finally, while explicit delensing of $B$-modes with these improved lensing maps will be demonstrated in upcoming work, we discuss the increase in correlation coefficient (and hence delensing efficiency) that the method can achieve with upcoming CMB data. }

\subsection{Wiener filtering}
Before turning to lensing reconstruction, we first demonstrate our Wiener filtering technique to accurately estimate the unlensed CMB from a simulated masked temperature CMB map. We convolved an input simulated lensed CMB sky with a beam of $7$-arcmin FWHM, and added homogeneous $35 \mu$K-arcmin noise. Using the exact input lensing potential and input unlensed spectrum, we obtained the reconstructed unlensed CMB modes $T^{\rm WF}$, up to $\ell \le 3500$. The mask was chosen arbitrarily as a piece of the public \planck\ 2015 analysis lensing mask, and we have further excised a band along all sides of the $600\: \rm{deg}^2$ patch so that the boundary of the unmasked patch is non-periodic. This leaves $f_{\rm patch} = 60\%$ of the patch unmasked.

The upper panels of Fig.~\ref{fig:filt} show a comparison of simulated lensed data and the estimated unlensed CMB ($T^{\rm WF}$) using the known input deflection field. The lower panels compare the residuals $T^{\rm WF} - T^{\rm input}$, using either the deflection-dependent Wiener filter ($\deflect = \nabla \phi^{\rm input},C^{\rm unl}_\ell$, right), or the usual quadratic estimate filter that we use for the initial estimate ($\deflect = 0,C^{\rm len}_\ell$, left). The former has small nearly-isotropic, near-uniform residuals set by the noise map, the latter shows the swirly patterns characteristic of lensing, with residuals of much larger amplitude.

It takes a couple of minutes on a modern laptop to reconstruct these modes, up to the point where the residual norm of the solution to the linear system of equations has decreased across the full patch by 5 orders of magnitude. At this point the values inside the masked regions have converged to sub-percent level.
\subsection{Lensing reconstruction on a masked patch}
\label{Sec:maskedrecon}

Next, we demonstrate lensing potential reconstruction from a simulated futuristic polarization-based experiment, including masking. We consider, on the same mask, the iterative reconstruction of the lensing potential from Stokes $Q,U$ simulated maps with noise level of $1.5 \cdot \sqrt{2}\: \mu$K-arcmin, and a beam of $3$ FWHM. This corresponds roughly to noise levels expected for the baseline, widefield CMB-S4 configuration. We use a vanishing fiducial gravitational wave amplitude, $C^{BB,\rm unl}_\ell = 0$, in which case the Wiener filter reconstructs the optimal unlensed $E$ mode map from $Q$ and $U$, assuming the $B$ mode is due to lensing and noise exclusively. We reconstruct multipoles of the $E$-mode map up to $\ell \leq 3500$, and the lensing potential over the same multipole range. The search converges just as well to higher multipoles, but there is little information there since the reconstruction becomes completely noise dominated. At no point do we apply low multipole cuts to the lensing potential, demonstrating that the iterative search can handle masking and the resulting large mean field adequately.

Convergence of the iterative search towards the optimal solution can be explicitly checked on all scales by monitoring the change in the posterior gradient and its components.
The upper panel of Fig.~\ref{fig:gplots} show the power spectra of the different pieces of the posterior gradient $g$ at the starting point of the iteration, for which the lensing reconstruction is given by the MV estimate.  These gradient maps are smooth across the mask, so to estimate the spectra we simply rescale naive spectra estimates by $1/{f_{\rm patch}}$ for the purpose of this figure. Assigning Fourier modes to $L$-bin according to $L = |\vecell| - 1/2$, resulting in $n_L \sim (2L + 1) V / 4\pi$ modes per bin, we build
\beq
\hat C^{\bg}_L = \frac 1 {f_{\rm patch}}\frac{1}{n_L}\sum_{ \vecell \textrm{ in } L \textrm{ bin }}|\bg_\vecell|^{2}.
\enq
Shown are the quadratic piece (blue), the mean field (orange) and the prior (green) spectrum. The total gradient spectrum is shown as the red line. For easier comparison, all gradients have been normalized with the quadratic estimator normalization $N^0_L$,  so the quantity plotted is actually
\beq
\lp N_L^0 \rp^2 \hat C^{\bg}_L,\quad \textrm{for } \bg = \bg^{\rm QD},\bg^{\rm MF},\bg^{\rm PR} \textrm{ and } \bg^{\rm tot}.
\enq
With this normalization, the reconstruction noise in the quadratic estimate before any iteration is $N^0_L$ itself, shown as the solid black line. On scales with large signal to noise ($N^0 \ll C_L^{\phi\phi}$) the inverse curvature is (to a crude approximation) $\sim N^0_L$. Thus, on these scales, $\lp N_L^0 \rp^2 C_L^{\bg \rm tot}$ is also roughly the spectrum of the Newton increment in Eq.~\ref{Newton} added to the potential estimate at the corresponding iteration step.

The quadratic part of the gradient is pure reconstruction noise at small scales and is cancelled by the prior gradient. On large scales, the quadratic piece is dominated by the mean-field contamination from the mask. Since the mask mean field is basically independent from $\hat \phi$, it will vary little from iteration to iteration, and improving the potential estimate (by reducing the total gradient) at low multipoles demands precise evaluation of this term. On the other hand, on intermediate scales we can see that $\bg^{\rm QD}$ is the dominant contribution to the total gradient at the start of the iterations. Hence, on these scales the lensing map can be improved without relying on cancellation of the mean field.

We used 511 simulations at each step to estimate the mean field. The upturn of the orange curve at $L \simeq 1000$ shows the onset of the MC noise dominated regime, where the mean-field estimate becomes pure MC noise. The dot-dashed black line shows an analytic prediction for the expected MC noise neglecting sky cuts, calculated with the tools from Appendix~\ref{appendix:MF}. The MC noise spectrum is smaller than $N^0_L$ by four orders of magnitude, so iterations should be able to reduce the gradient amplitude by a similar amount.  On large scales, the MC noise stays smaller than the prior gradient, and thus the iterative procedure will have exhausted information from the data before it hits the MC noise floor. However, we will see below that the isotropic prediction for the MC noise is inaccurate on large scales, where there is a substantial contribution from sky cuts.
A small contribution also comes from the deflection field. It is possible to predict this contribution perturbatively, since the mean field just follows the spatial distribution of the deflection field at each step: see Appendix \ref{appendix:MF}. This contribution from the deflection is comparatively larger for temperature reconstruction, and also for temperature in combination with polarization.

We start the posterior inverse curvature $H^0$ with the isotropic estimate
\be
H^0_L = \lp \frac{1}{N^{0,\rm unl}_L} + \frac{1}{C^{\phi\phi}_L}\rp^{-1}.
\enq
The second term is the prior curvature, and for the first term (the likelihood curvature) we used the unlensed weights. The choice of initial curvature is not  critical as long as the BFGS scheme is used to update it. However, using the lensed weights can lead to the algorithm taking steps that are too large and hence give poor convergence, so we use unlensed weights instead (which gives a curvature that is slightly too large, but works well): steps that are too large should be avoided, as the search relies on the displacement being invertible at each step. More optimal curvature estimates might be built using partially lensed weights.

Convergence is acceptably quick, and after $\sim 9$ iterations the bulk of the improvement has been gained, with the gradient spectrum reduced by $2$--$3$ orders of magnitude. Only small variations on the large-scale modes are visible in the deflection maps after this point. The lower panel of Fig~\ref{fig:gplots} shows the spectra after 20 iterations. The purple curve show an empirical estimate of the mean-field MC noise. This is calculated by splitting our set of simulations into two independent sets of size $N_1$ and $N_2$ with corresponding mean-field prediction $\bg^{\rm{MF}_1}$ and $\bg^{\rm{MF}_2}$, and building
\beq
\hat C_L^{\rm MF} = \frac{N_1N_2}{\lp N_1 + N_2\rp^2} \hat C_L^{\rm{MF}_1 - \rm{MF}_2}.
\enq
This is much larger than the analytic isotropic prediction on large scales because of the mask contribution to the MC noise. The total gradient closely follows the MC noise curve, and no further improvement can be achieved after the 20 iterations. At intermediate scales, the quadratic gradient is visibly much reduced and is now in equilibrium with the prior.

Finally, we show the reconstructed lensing map in Fig.~\ref{fig:QUrec}. From left to right in the top row we show the input lensing map, the quadratic estimate, and the converged iterative solution. Here we plot the displacement-like but isotropic spin-0 transforms
\beq
\label{dx}
d(\vecx) \equiv\frac 1 {\sqrt{V}} \sum_{\vecL}L\: \hat\phi_\vecL \:e^{i\vecL \cdot \vecx}.
\enq
The reconstruction is visibly improved, both by large-scale modes filling in the masked regions and by the presence of finer-grained structure well inside the patch. The bottom row of Fig.~\ref{fig:QUrec} shows a zoom in of the central area, showing instead the lensing convergence where the improvement on small scales is more clearly visible.

\subsection{Delensing efficiency}\label{subsec:eps}
\revision{How can our lensing reconstruction method help with measurement of primordial tensor modes? In the absence of a fiducial non-vanishing $C_\ell^{BB}$, for which there is at present no preferred choice, no delensed $B$-mode map is directly produced by the algorithm as the prior sets it to zero. However, it is well known that the lensing map can be used to remove some of the lensing signal in the observed $B$-mode map. Reduction of lensing signal in the $B$-mode map will result in some degree of improvement on tensor constraints, since the lensing $B$ modes act as a source of noise for any primordial signal.}

\begin{figure}
 \includegraphics[width = 0.5\textwidth]{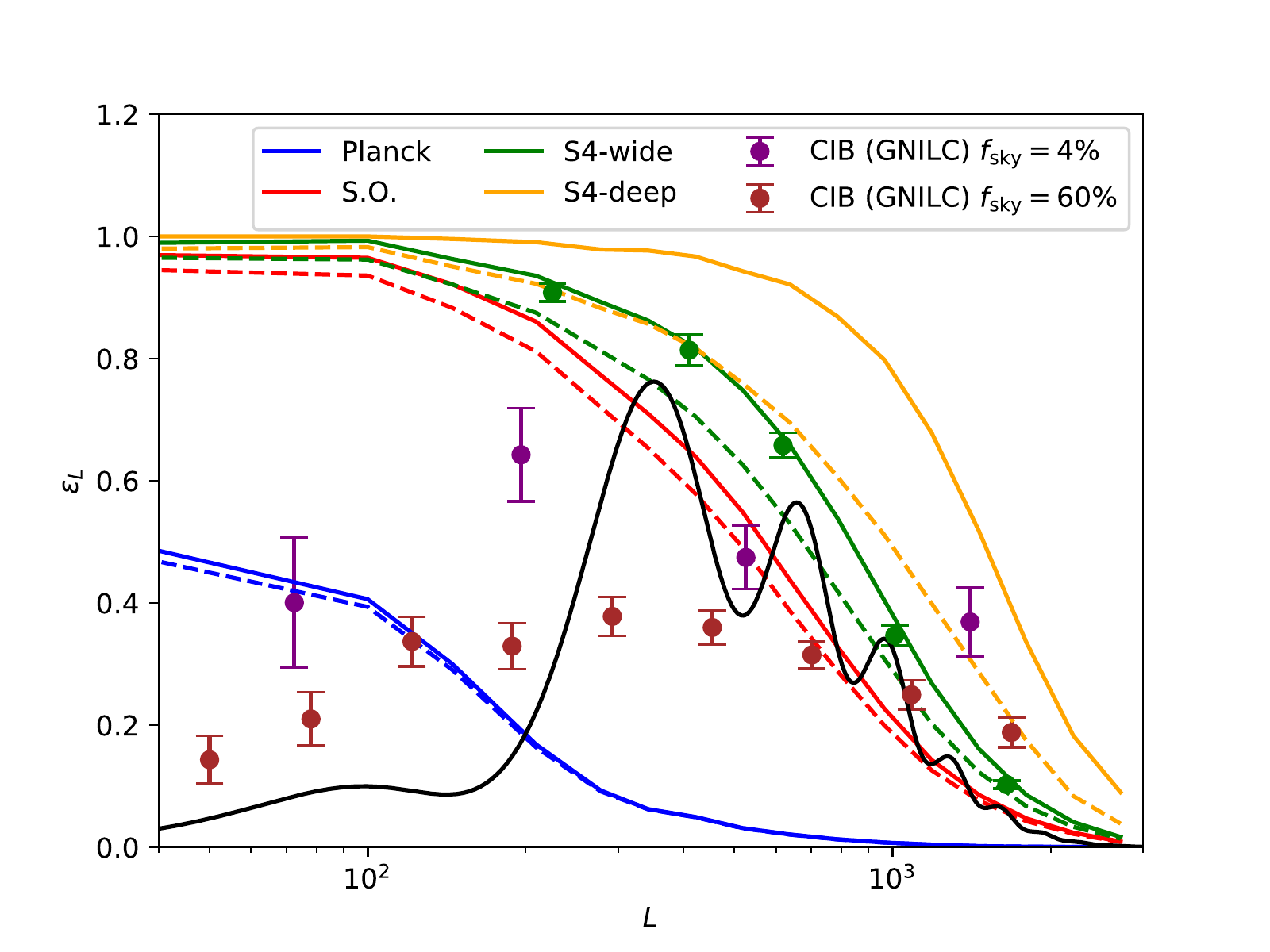}
 \caption{\label{fig:epsilons}The delensing efficiencies as function of lensing $L$ multipole with either the quadratic estimator (dashed coloured lines) or the iterative solution (solid coloured lines), for current and futuristic noise levels.
 The curves were obtained from 128 simulated spectra and cross-spectra of the input lensing with idealized quadratic and iterative reconstructions. The green points show for comparison an estimate of the efficiency from the reconstruction on the masked sky described in Sec.~\ref{Sec:maskedrecon}, which has identical noise level to the corresponding green curve. Only data far away from the mask edges were used to produce these points.
  Also shown are estimates of the efficiency using CIB maps, obtained as discussed in the main text from the public \planck\ \GNILC\ maps at $545\,{\rm GHz}$. The brown points show the efficiency obtained from $60\%$ of the sky, while the purple points were obtained on the cleanest (according to the \GNILC\ dust map) $4\%$ of the sky.
  The black line shows the contribution per log-multipole bin of  $C_L^{\phi\phi}$ to the total $B$-mode power on the scales relevant for a primordial $B$-mode measurement\revision{, see Eq.~\ref{blackline}}. The average of the coloured curves weighted by the black line gives the approximate delensing efficiency relevant to each observation. The residual lensing $B$ power is listed in Table~\ref{table:eps} together with the noise levels and expected improvement on tensor-to-scalar ratio constraints.}
 \end{figure}

 Delensing of $B$-mode polarization has recently  been demonstrated on \planck\ data by remapping the Stokes maps~\cite{Carron:2017vfg}, and by the SPT team \cite{Manzotti:2017net} using a template subtraction method.
 In both cases, the expected reduction of lensing-like power is approximately set by the squared cross-correlation coefficient of the measured lensing map to the true lensing map, which we call the delensing efficiency:
 \beq\label{epsilon}
 \epsilon_L \equiv \frac{\lp C_L^{\hat \phi \phi}\rp^2}{C_L^{\phi\phi}C^{\hat \phi \hat \phi}_L}.
 \enq
Fig.~\ref{fig:epsilons} shows this cross-correlation coefficient for simulated reconstructions. We built these curves using 128 idealized simulations, with homogeneous input noise maps and no sky cuts.
In this case, the deflection field is the only source of mean field at each step, which is sufficiently well described by the perturbative predictions derived and discussed in the Appendix~\ref{appendix:MF}. In all cases, we have considered joint temperature and polarization (MV) reconstruction, with sharp multipole cuts $10 < \ell \le 3000$. For the \planck\ curve, we cut at $2048$, following the public analysis~\cite{Ade:2015zua}. Shown are the delensing efficiencies expected for the MV quadratic estimator (dashed colour) and the iterated, converged solution (solid). Besides \planck, we also show curves for the \SO\footnote{\url{www.simonsobservatory.org}}, and two distinct CMB-S4-like configurations: one for a wide but shallow coverage, and one for a deep survey with sensitivity increased by a factor of about four. 
The assumed beam and noise levels are shown in  Table~\ref{table:eps}, ignoring all experimental complications, and are not meant to be necessarily very accurate representations of the experiment label. We have used Gaussian beams of 3 arcmin FWHM in all cases, again with the exception of \planck\  where we used 6.5 arcmin. The S4-wide configuration is identical to those of the optimal reconstruction performed on the masked sky in Sec.~\ref{Sec:maskedrecon}. We show an estimate of the efficiency of this reconstruction as the green data points. The points are obtained from Eq.~\ref{epsilon}, using pseudo-$C_\vecL$ estimates after enlarging the mask conservatively near the mask boundaries, leaving $f_{\rm patch} \sim 35\%$, in order to avoid any edge effects. The points stand in very good agreement to expectations, demonstrating that masking does not substantially affect the reconstruction quality away from the mask boundaries.

Fig.~\ref{fig:epsilons} also shows the delensing efficiency reachable using the publicly available\footnote{\url {http://pla.esac.esa.int}} \GNILC\ Cosmic Infrared Background (CIB) map \cite{Aghanim:2016pcc} as an external tracer of the lensing map (purple points).
We used the \GNILC\ reconstruction at $545$ GHz, covering $60\%$ of the sky, and \planck\ 2015 lensing potential map to build these points. We estimated the CIB auto spectrum and the lensing-CIB cross-spectrum on the union of their released masks after apodization on a scale of $12\,\arcmin$, deconvolving the pseudo-$C_\ell$ estimates from the mask coupling matrix. We show the CIB efficiency estimate
\beq
\hat \epsilon^{\rm CIB}_L \equiv \frac{\lp \hat C_L^{\rm CIB \hat \phi}\rp^2}{\hat C^{\rm CIB\,\rm CIB}_LC_L^{\phi\phi,\rm fid}}
\enq
as the brown data points. The fiducial lensing spectrum is based on the \planck\ 2015 cosmology. This estimate is justified in so far as the lensing map is an unbiased tracer of the true lensing, and in the absence of spurious cross-correlation between the two maps. Comparison to previous works on CIB delensing (Fig.~2 of Ref.~\cite{Larsen:2016wpa}, using \planck\ cleaned $545$ GHz map, and Fig.~1 of Ref.~\cite{Manzotti:2017net} from the SPT team, using \herschel\ 500 $\mu m$ map) shows good consistency\footnote{Note that both references show the cross-correlation coefficient $\rho_L$, while we show the efficiency $\rho_L^2$} over the relevant scales.

On this large sky fraction ($60\%$), contamination by galactic dust may reduce somewhat the cross-correlation to the lensing in the GNILC CIB map, and the brown points are slightly lower than expected for clean maps \cite{Ade:2013aro,Sherwin:2015baa}. Larger CIB efficiencies might be possible on cleaner regions of the sky, or with improved dust cleaning from future observations. For comparison we also show as the purple data points the efficiency on a smaller but cleaner area, using a mask built by thresholding the GNILC dust map at 545 GHz, keeping only $4\%$ of the sky unmasked.

	
 \revision{Not all multipoles are equally important for the purpose of $B$-mode delensing. To a good approximation, the $B$ power $C_\ell^B$ depends linearly on the lensing deflection spectrum, hence we may write the delensed $B$-mode power as
 \beq\label{blackline}
 C^{B,\rm{delens}}_\ell \sim \sum_{L}(1 - \epsilon_L) \frac{\partial C^{B}_\ell}{\partial \ln C_L^{\phi\phi}} .
 \enq
 If the very-low $\ell$ reionization peak cannot be probed or is discarded, the tensor-mode recombination peak (at roughly $40 \leq \ell \leq 100$)
 determines the scale where delensing is most important.
The black line on Fig.~\ref{fig:epsilons} shows $L d C^{BB}_\ell/ d\ln C_L^{\phi\phi}$, after averaging over this $\ell$-multipole range, and normalized such that it $L$-integrates to unity. By construction, weighting the efficiency curves on this figure against this line gives the delensing efficiency relevant for primordial $B$ modes around the recombination peak.}

Effective residual delensed $B$-mode noise amplitudes are listed on the second set of rows of Table~\ref{table:eps}.  The delensed $B$-mode lensing power is calculated from the unlensed $E$ spectrum and a reduced lensing spectrum given by
\beq\label{delens}
 C_L^{\phi\phi,\rm delens} = (1 - \epsilon_L)C^{\phi\phi}_L,
\enq
 where the efficiencies are those shown in Fig.~\ref{fig:epsilons}. The numbers in the table are the mean power over $40 \le \ell \le 100$.
The partially lensed spectra as well as the coupling matrix $d C^{BB}_\ell/ d\ln C_L^{\phi\phi}$ are obtained with the Python \CAMB\ package\footnote{\href{http://camb.readthedocs.io/en/latest/correlations.html}{camb.readthedocs.io}}. The \planck\ number matches well the result of the $B$-mode delensing analysis performed by Ref.~\cite{Carron:2017vfg} on data. We also give predictions for the iterated solution. These predictions are obtained following Ref.~\cite{Smith:2010gu} by iteratively producing delensed power spectra and MV reconstruction noises $N_L^{0}$, using at each step the reduced lensing power in Eq.~\ref{delens} with efficiencies
 \beq
 \epsilon_L = \frac{C_L^{\phi\phi}}{C_L^{\phi\phi} + N_L^{0}}
 \enq
 to calculate the partially delensed $B$-mode power used when calculating $N_L^{0}$ for the next step.
 The predictions stand in excellent agreement with our simulated reconstructions.

 Table~\ref{table:eps} also shows the expected delensing improvement of constraints on the tensor-to-scalar ratio $r$, comparing results using the iterative lensing estimator to those using the quadratic estimator. These are calculated from a toy $r$ estimator variance estimate, assuming $r = 0$,
\beq
\frac 1 {\sigma^{2}(\hat r)} = \frac{f_{\rm sky}}2\sum_{\ell \geq 40} (2\ell + 1)\lp \frac{C_\ell^{B\:\rm {tensor, r = 1}}}{C_\ell^{B\:\rm delens}+C_\ell^{B\:\rm noise}} \rp^2.
 \enq
The delensed $B$ power is the one calculated according to Eq.~\ref{delens}.
More realistic forecasts, for example including foreground cleaning, are well beyond the scope of this paper, but we note that these ratios stand in agreement with expectations~\cite{Abazajian:2016yjj}.


\begin{table}\caption{The first three rows give the temperature and polarization noise levels and beam width input to the simulations used in Sec.~\ref{subsec:eps} to obtain the delensing efficiencies shown on Fig.~\ref{fig:epsilons}. The next two rows show the effective $B$-mode lensing power achievable on the scales relevant for primordial $B$-mode measurement, using the quadratic (MV) estimator and the iterated solution respectively. The latter results accurately match predictions using an iterated Gaussian noise level as described in the main text, shown on the sixth row. The last two rows show the fractional improvement on the error bar of the tensor to scalar ratio $r$ (compared to the case of no delensing, assuming $r = 0$). These results are for the idealized case of no foreground or mean-field contamination, and show some sensitivity to the largest multipole $L$ that can be delensed. Ratios calculated using a lens multipole cut at $L_{\rm{min}} = 100$ instead of $L_{\rm{min}} = 40$ are shown in parentheses.}
\begin{tabular}{lccccc}
\label{table:eps}
 & Planck & S.O. & S4-wide & S4-deep \\
\hline
$N_{\rm lev}^T$ / ($\mu $K arcmin) & 35 & 3.0 & 1.5 & 0.38 \\
$N_{\rm lev}^P$ / ($\mu $K arcmin) & 55 & 4.2 & 2.1 & 0.53 \\

Beam FWHM $/$ (arcmin) & 6.5 & 3.0 & 3.0 & 3.0 \\
\hline
$B^{\rm len}$  / ($\mu $K arcmin)  (quadratic) & 4.5 & 3.2 & 2.8 & 2.3 \\
$B^{\rm len}$  / ($\mu $K arcmin)  (iterative) & 4.5 & 3.0 & 2.4 & 1.4 \\
pred. (iter. $N^0_L$) & 4.5 & 3.0 & 2.4 & 1.4 \\
\hline
$\sigma(r)$ impr. (quadr.) & 1.0 & 1.7(1.5) & 2.5(2.2) & 4.1(3.3) \\
$\sigma(r)$ impr. (iter.) & 1.0 & 1.8(1.6) & 3.1(2.6) & 10.6(6.3) \\
\hline
\end{tabular}
\end{table}
\section{Summary}\label{sec:summary}
We presented an iterative method for CMB lensing reconstruction, and showed that for future high-sensitivity observations it can produce substantially better results than the quadratic estimator. Even with non-trivial masking, the method remains numerically tractable and produces results in agreement with naive expectations.
For low noise levels the large-scale lensing modes are all reconstructed with high signal to noise, even by the quadratic estimator, so the cosmological information is limited by cosmic variance. The main information gain from the iterative estimator comes on smaller scales where the quadratic estimator reconstruction noise starts to be substantial. However, for $B$-mode delensing, even small errors on the reconstruction of the large-scale lensing realization can lead to residual lensing $B$-mode power, so the improvement in signal to noise is important on all scales.

The algorithm works by extracting residual lensing from optimally reconstructed unlensed CMB maps. As such, it produces both estimates of the lensing potential and the delensed CMB maps. Note, however, that the method does not directly produce a delensed $B$-mode map, unless a prior spectrum is adopted for the unlensed $B$-mode spectrum. Nevertheless, the resulting deflection estimate, alone or in combination with the delensed $E$ map, may be used to delens the observed polarization map, giving improved delensing efficiency compared to using a quadratic estimator reconstruction.

The algorithm maximizes the posterior probability for the lensing potential, assuming Gaussianity of the unlensed maps and noise. \revision{A solution to this same problem was first attempted in Refs.~\cite{Hirata:2002jy,Hirata:2003ka}. These references, working in the absence of non-ideal effects, introduced several approximations to reduce the computational burden, avoiding in particular the anisotropic inverse variance filtering step. For similar reasons involving the difficulty of a global analysis, Ref.~\cite{Anderes:2010fq} introduced a local likelihood reconstruction method, where the lensing map is approximated as quadratic in small neighborhoods, and large wavelengths are ignored. We have demonstrated how a conjugate gradient inversion can handle the global inversion very efficiently, and, crucially, can also successfully be applied in the presence of sky cuts and other realistic non-idealities. Our solution is the first that does not rely on approximations once the fiducial ingredients of the likelihood and prior have been chosen.}
This means that, given enough computational resources, the resulting lensing potential map is optimal and cannot be improved upon.

In practice, one limiting factor is the mean-field calculation. Unless some approximation is used, the mean field is calculated with a finite number of simulations, and this sets a Monte-Carlo noise floor that cannot be improved upon by further iterations. However, we demonstrated that for realistic situations reconstructions can be successfully performed on masked data for current and next-generation CMB experiments. \revision{We also showed how a perturbative approximation to the mean field is adequate in the absence of non-ideal effects, allowing very fast iterative reconstructions in this case.} We expect the methods and codes described and tested here to be useful for the planning and execution of future CMB lensing analyses.

The modular, fully parallelized pipeline (using MPI) is written in Python, internally calling parts written in C, and/or sending these to a GPU device using the pyCUDA interface \citep{pyCUDA2012}. \revision{The flat-sky code is publicly available\footnote{\url{https://github.com/carronj/LensIt}}.}
We also described the curved-sky  algorithm; this will be tested on data and reported elsewhere.

Our successful exploration of the lensed CMB likelihood suggests several interesting possibilities for future investigation and improvement. The iterative estimate takes as an input the fiducial unlensed CMB spectra, which we have taken to include no primordial $B$ modes so that our posterior (MAP) estimate of the unlensed $B$ modes is exactly zero. This prevents us directly obtaining an optimal measurement of a delensed gravitational wave signal, which must be obtained afterwards using a more standard template subtraction or point remapping method. By allowing for non-zero unlensed $B$ modes in the prior, we could also allow direct joint estimation of the lensing together with the primordial signal. Exactly how best to do this, given the unknown amplitude of the primordial signal and complications with delensing biases, is worth careful future consideration.
 Another important future direction is to go beyond estimation of the lensing map to also provide optimal lensing power spectrum estimates (and estimates of the delensed CMB power spectra). Within the maximum posterior density framework, building a posterior density for the lensing power spectrum formally requires an intractable marginalization over the deflection field, though approximations can certainly be built~\cite{Hirata:2003ka} and are worth further study. Finally, this also opens exciting prospects for cluster CMB lensing~\cite{Baxter:2014frs,Madhavacheril:2014slf}, by allowing non-parametric cluster mass profile measurements from the full likelihood. 

\begin{acknowledgments}
We thank Duncan Hanson for making his \textit{qcinv} code publicly available, on which our filtering code is partly based.
The research leading to these results has received funding from the European
Research Council under the European Union's Seventh Framework Programme
(FP/2007-2013) / ERC Grant Agreement No. [616170]. This research used resources of the National Energy Research Scientific Computing Center, a DOE Office of Science User Facility supported by the Office of Science of the U.S. Department of Energy under Contract No. DE-AC02-05CH11231. Part of this paper is based observations obtained with Planck, an ESA science mission with instruments and contributions directly funded by ESA Member States, NASA, and Canada.
\end{acknowledgments}
\clearpage
\newpage
\onecolumngrid
\appendix
\section{Curved sky gradients} \label{appendix:curved}
\newcommand{\Ylm}[1]{\left._{#1}Y_{lm}(\hn)\right.}
\newcommand{\Ylms}[1]{\left._{#1}Y_{lm}^*(\hn)\right.}
\newcommand{\Res}[1]{\left._{#1}\textrm{Res}\right.}

\newcommand{\gMAP}[1]{\left._{#1}\nabla X\right.}
\newcommand{\geo}{+}
\newcommand{\kls}[2]{\sqrt{#1\lp(#1+1\rp - #2\lp#2-1\rp}}
\begin{figure}
\includegraphics[width = 0.5\textwidth]{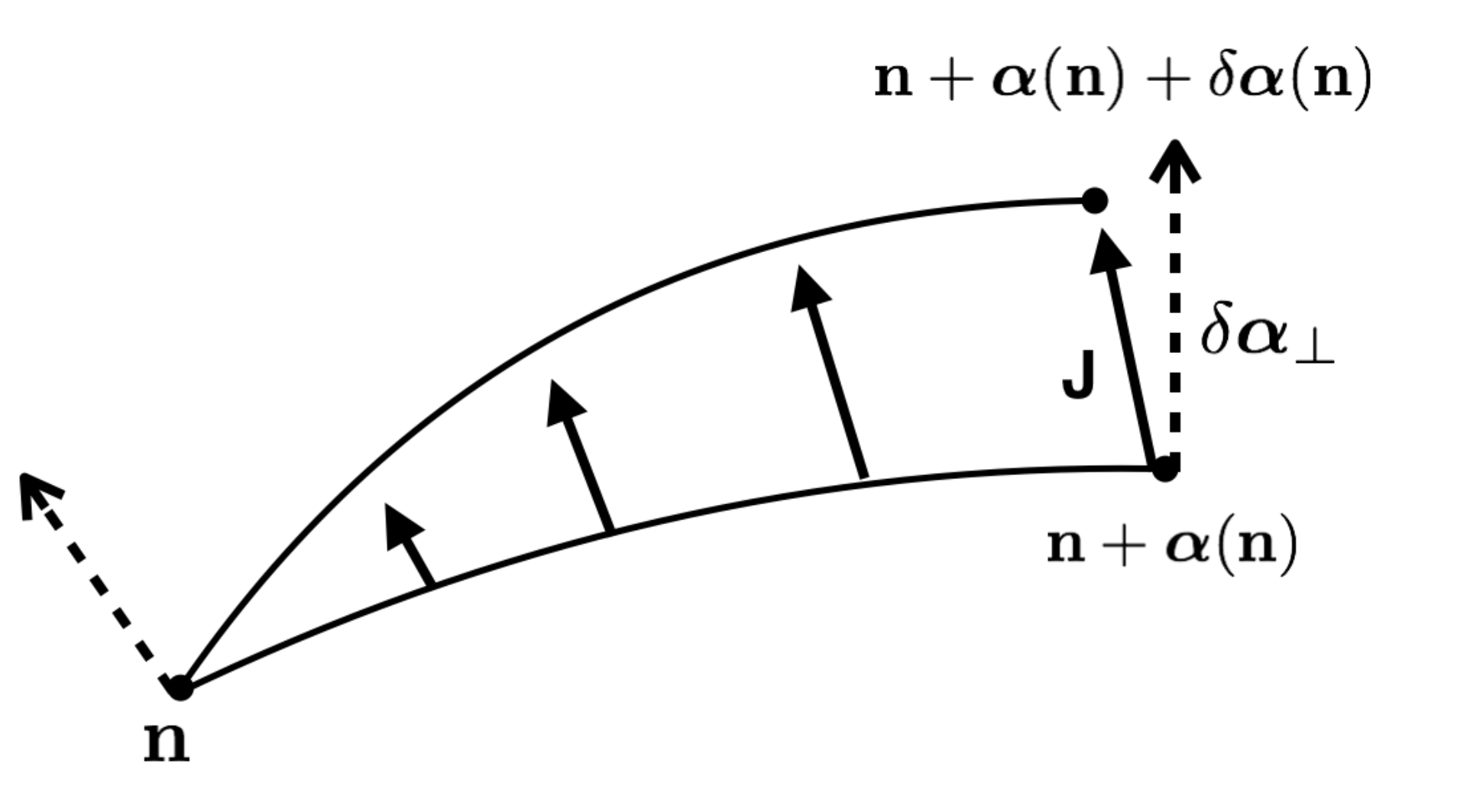}
\label{fig:geo}
\caption{Schematic aid to the curved-sky lensed CMB likelihood gradient calculation. The variation of a lensed tensor $\mathcal T (\hn + \deflect(\hn))$ with respect to the deflection must be calculated at the deflected position $\hn + \deflect(\hn)$. This point is defined by following the geodesic from $\hn$ in the direction $\deflect$ for length $|\deflect|$. A variation $\delta \deflect(\hn)$ in the deflection vector at $\hn$ shifts the geodesic slightly. The exact first order change in position is given by the vector field $\bf J$ (solid arrows), proportional to $\delta \deflect$ infinitesimally close to $\hn$  and evolving along the original geodesic according to the Jacobi equations set by the curvature tensor. For simplicity, we instead evaluate the variation using the parallel-transported $\delta\deflect$ (dashed arrows) instead of $\bf J$. This neglects the focussing effect of the sphere curvature, slightly overestimating the geodesic deviation. The relative error in the gradient normal component is quadratic in the deflection angle and equal to $(1- \sin \normdeflect / \normdeflect) \sim 6\cdot10^{-8}$ for $~2 $ arcmin deflections. This is completely negligible and of similar order as the small-angle approximation and neglected physical effects such as polarization rotation~\cite{Lewis:2017}: it is safe to neglect sky curvature on the scale of the deflection angles.}
\end{figure}
We give here the curved sky likelihood gradients, analogous to the flat sky version given in the main text. We first state the results and describe the implementation, and then provide a derivation.
This requires only repeated use of the gradient and curl decomposition of a complex spin $s$ field, which are readily available in widespread packages.
With real part $\mathcal R$ and imaginary part $\mathcal I$, we may write $_{\pm |s|}f$
as
\beq
 \left._{\pm |s|}\right. f(\hn) = \left( \mathcal R(\hn) \pm i \mathcal I(\hn)\right).
\enq
Then the gradient curl component are defined as ($|s| > 0$)
\beq \label{GClm}
\int d^2n \Ylms{\pm |s|} \left._{\pm |s|} f(\hn)\right. \equiv  - \lp \pm 1 \rp^{|s|} \lp G_{lm} \pm i C_{lm} \rp,\quad \leftrightarrow \left._{\pm |s|}\right. f(\hn)  =- (\pm 1)^{|s|} \sum_{lm} \lp G_{lm} \pm i C_{lm} \rp \Ylm{\pm |s|}.
\enq
We follow here for convenience the sign conventions adopted e.g. by the relevant spin harmonic transform routines of the widespread \HEALpix\ package~\cite{Gorski:2004by}. The definitions of the spin harmonics follow e.g.
Ref.~\cite{Lewis:2001hp}.
The $E,B$ decomposition of the spin-$2$ polarization field is
\beq
 \left._{\pm 2}P\right.(\hn) \equiv Q (\hn) \pm iU(\hn) = - \sum_{l m}\lp E_{lm} \pm iB_{lm} \rp \Ylm{\pm 2}.
\enq
We give the results for the joint $T,Q,U$ analysis, the restriction to temperature only or polarization only is straightforward.


We aim to obtain the likelihood gradients with respect to the displacement modes, in analogy to the flat sky derivation in the main text. We only need to derive the quadratic part of the gradient: the mean field is, as before, its average, and the Gaussian prior is straightforward. We define it by projecting the gradient vector onto the spin basis $\ve_{\pm}$. This basis is associated to the cartesian orthonormal frame $\ve_1,\ve_2$ orthogonal to $\vn$, given by $\ve_{\pm} \equiv \ve_1 \pm i\ve_2$. The spin $\pm1$ quadratic gradient is then defined as
\beq
\label{bg}
_{\pm 1} g^{\rm QD}(\hn) \equiv  e_{\pm}^a \frac{\delta }{\delta \normdeflect^a(\hn)}\lb\frac 12 \Stdat\cdot \Cov_\deflect^{-1} \Stdat \rb.
\enq
The unnormalized potential ($\phi$) and curl potential ($\Omega$) quadratic estimators (as could have been obtained directly by taking gradients with respect to $\phi$ and $\Omega$)
are then simply given by the harmonic expansion of $_{\pm 1}g$:
\beq
\label{g1}
_{\pm 1}g^{\rm QD}(\hn) = -(\pm 1)\sum_{LM} \lp\frac{ \phi^{\rm QD}_{LM} \pm i \Omega_{LM}^{\rm QD}}{\sqrt{L(L + 1)}}\rp  \left. _{\pm 1}Y_{LM}(\hn)\right. .
\enq

Postponing the derivation, the end result is as follows: with $\hn$ and $\hn'$ the undeflected and deflected points,
\beq\label{g1result}
_{1} g^{\rm QD}(\hn) = -\sum_{s = 0,\pm 2} \Res {-s}(\hn) [\eth {}_sX^{\rm WF}](\hn').
\enq
In this equation, the left leg of the quadratic product is the inverse noise weighted residual
\beq
\label{Res}
\Res{}(\hn) \equiv \lb B^\dagger \Cov_\deflect^{-1} \Stdat \rb(\hn) = \lb B^\dagger N^{-1}\lp   \Stdat- B \Dop X^{\rm WF}\rp\rb(\hn),
\enq
and the right leg is given by deflected gradients of the Wiener-filtered maps. Explicitly,
\beq
\label{gMAP}
\begin{split}
[\eth {}_0X^{\rm WF}](\hn') &= \sum_{lm}\sqrt{l (l + 1)} T^{\rm WF}_{lm} \left._1Y_{lm}(\hn')\right. \\
[\eth {}_{-2}X^{\rm WF}](\hn') &= -\sum_{lm} \sqrt{(l+2)(l - 1)} \lb E^{\rm WF}_{lm}-  i B^{\rm WF}_{lm}\rb\left._{-1}Y_{lm}(\hn')\right.\\
[\eth {}_2X^{\rm WF}](\hn') &=-\sum_{lm} \sqrt{(l-2)(l+ 3)} \lb E^{\rm WF}_{lm}+  i B^{\rm WF}_{lm}\rb \left._3Y_{lm}(\hn')\right. .\\
\end{split}
\enq
The only difference between Eq.~\ref{g1result} and traditional position-space curved-sky implementation of the quadratic estimator (as stated above, without the $N_0$ normalization) are the use of the unlensed spectra instead of the lensed spectra in producing $X^{\rm WF}$, together with the presence of the deflection operations, both in the filter and explicitly in Eq.~\ref{g1result}.


We now justify Eq.~\ref{g1result}. The model for the observed signal with noise $n$ is
\beq
\Stdat = B \Dop X^{\rm unl} + n ,
\enq
where on the curved sky the operator $\Dop$ sends the unlensed $T,E$ and $B$ CMB modes to the deflected Stokes map with definite spin $0,\pm 2$ (i.e. $_{\pm 2}P$ and not $Q,U$):
\beq
\label{Dlm}
D_{lm}(\hn)=  \begin{pmatrix} _0Y_{lm} &0  & 0 \\
		0   &  - _2Y_{lm} &  -i _2Y_{lm} \\
		0   &  -_{-2}Y_{lm} & i _{-2}Y_{lm}\ \\
\end{pmatrix}(\hn').
\enq
Similarly, $B$ projects the spin maps $T,_{\pm 2}P$ to the observed $\Stdat = T^{\rm dat}, Q^{\rm dat}$ and $U^{\rm dat}$.
From the definitions given in Eqs.~\ref{bg} and \ref{Res}, we have
\beq
\begin{split}
 _{1} g^{\rm QD}(\hn) &={\Stdat}^\dag\Cov_\deflect^{-1}  B \lb -e_{+}^a \frac{\delta D }{\delta \normdeflect^a(\hn)} \rb C^{\rm unl} D^\dagger B^\dagger \Cov_\deflect^{-1} \Stdat \\
 &=  \int d^2n'\sum_{s = 0,\pm 2} \Res {-s} (\hn') \lb-e_{+}^a  \frac{\delta D }{\delta \normdeflect^a(\hn)} X^{\rm WF}  \rb_{s+1}(\hn').
\end{split}
\enq
On the second line we used the spin $s$ as the index for the different components of the residual and gradient maps.
How to make sense of and evaluate the variations of $D$? From Eq.~\ref{Dlm}, all elements are spin-weighted harmonics at the deflected position, hence we need to understand how this position changes under a variation of the deflection. The geometry is sketched on Fig.~\ref{fig:geo}. On the curved sky, the notation $\hn \rightarrow \hn + \deflect(\hn)$ indicates displacement of length $|\deflect(\hn)|$ from $\hn$ along the geodesic in direction $\deflect$. The polarization axes are parallel transported along the geodesic, leading to some small change in $Q,U$ from the resulting misalignment with the coordinate vectors at the new point~\citep{Challinor:2002cd,Lewis:2005tp}. Varying $\deflect(\hn)$ by a small amount $\delta\deflect(\hn)$ give rises to a slightly different geodesic. On the flat sky, the end separation vector between the points will be $\delta\deflect(\hn)$, but this is not so on the curved sphere. Since $\alpha$ is typically a few arcminutes, the difference is very small, and we will neglect it. We now justify this, by deriving the exact but less practical result.

On any Riemannian manifold, the separation between the geodesics is described by the Jacobi vector $\bf J$. $\bf J$ is initially zero, has initial velocity $\delta\deflect(\hn)$, and its acceleration is set by the Riemann curvature tensor through the Jacobi equations. The positive curvature of the sphere will reduce the separation vector compared to the flat sky. The covariant, first order change in a tensor $\mathcal T$ on the manifold is
\beq
(\delta \mathcal T)(\hn') = (J^a\nabla_a \mathcal T)(\hn')
\enq
\newcommand{\epar}{\boldsymbol{e_{\parallel}}}
\newcommand{\eperp}{\boldsymbol{e_{\perp}}}

On the sphere, using a parallel orthonormal frame, with one vector $\epar$ initially aligned with $\deflect(\hn)$, $\bf J$ is given by
\beq
\boldsymbol{J}(\hn') = \delta \normdeflect^{\parallel}(\hn)\epar(\hn') + \frac{\sin \normdeflect(\hn)}{\normdeflect(\hn)} \delta \normdeflect^{\perp}(\hn) \eperp(\hn'),
\enq
Hence, in this frame,
\beq
(\delta \mathcal T)(\hn') =\delta \normdeflect_{\parallel}(\hn) \nabla^\parallel \mathcal T(\hn')  + \frac{\sin \normdeflect(\hn)}{\normdeflect(\hn)} \delta \normdeflect_{\perp}(\hn)\nabla^\perp \mathcal T(\hn') .
\label{deltaT}
\enq
This differs by $\sin \alpha / \alpha$ in the perpendicular component from the approximation that we use, where instead we use the parallel-transported $\delta\deflect(\hn)$ (and not $\delta\deflect(\hn')$),
\beq
\frac{\delta \mathcal T (\hn')}{\delta \normdeflect^a(\hn)} \approx \delta^{D}(\hn + \deflect(\hn) -\hn')\nabla_a \mathcal T(\hn').
\enq
 This is extremely accurate, since $1-\sin \alpha / \alpha \sim 10^{-7} $ for $\sim 2$ arcminutes deflections.
With this approximation, we can make use of the spin lowering and raising form of covariant derivatives for spin weighted functions~\cite{Lewis:2001hp,Challinor:2002cd}.
The equivalent of Eq.~\eqref{deltaT} for spin-weight quantities $_s\mathcal T$ is
\beq
e_{+}^a \frac{\delta  {}_s \mathcal T(\hn')}{\delta \normdeflect^a(\hn)}
= 2\frac{\delta  {}_s\mathcal T(\hn')}{\delta {}_{-1} \normdeflect(\hn)}
\approx -\delta^{D}(\hn + \deflect(\hn) -\hn')\eth {}_s \mathcal T(\hn').
\enq

Using repeatedly
\beq
\eth _sY_{lm} = \sqrt{l(l+1) - s(s + 1)} _{s + 1}Y_{lm}
\enq
 on all the $D$ matrix entries of Eq.~\eqref{Dlm} gives the result in Eqs.~\ref{g1result} and \ref{gMAP}.

\section{Mean field} \label{appendix:MF}
 \begin{figure}
 \includegraphics[width = 0.5\textwidth]{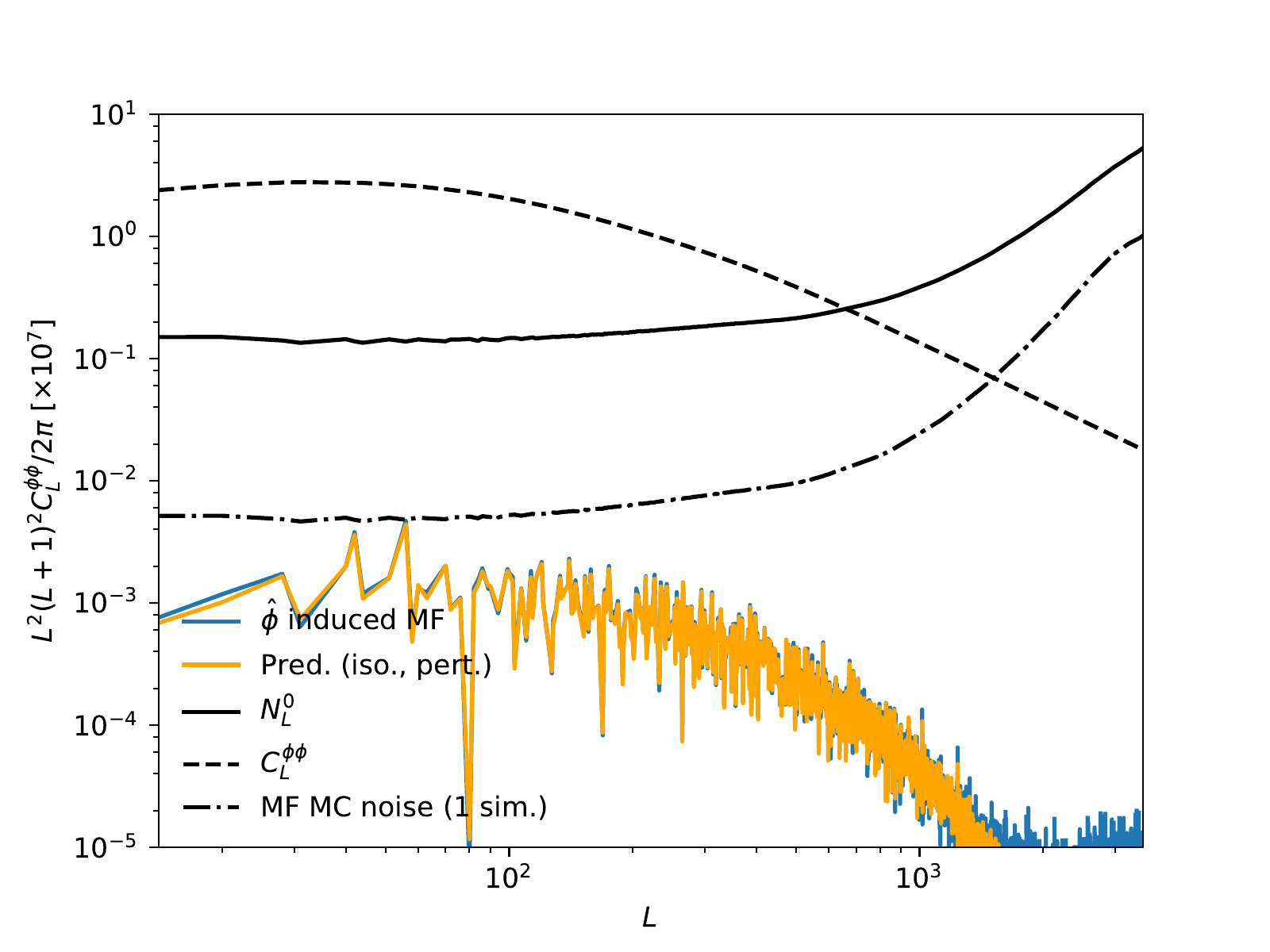}
\caption{\label{fig:MFpred}The expected contribution of the deflection-induced mean field for the first step of iterative reconstruction from polarization with noise level of $1.5 \cdot \sqrt 2 \mu$K-arcmin and no sky cut. The prediction (orange line) is calculated using an input deflection with spectrum equal to that expected for the first iteration estimate of the maximum a posteriori solution (i.e. the spectrum of the Wiener-filtered deflection, Eq.~\eqref{WienerPhiSpectrum}). Also shown as the (barely visible) blue curve is the measured mean field, with MC noise visible on small scales. The accuracy of the prediction is at least percent-level. Both curves were normalized by $N^0$, as the gradients on Fig~\ref{fig:gplots}, and can be directly compared. This shows that the $\phi$-induced mean field plays very little role in this configuration on the first iteration, where the bulk of the reconstruction improvement is performed. The dotted-dashed black line shows (with the same normalization) the single-simulation Monte-Carlo noise of the mean-field estimator used in this work, Eq.~\ref{Acc}. For this configuration, it improves upon the naive $N_0$ noise (solid black) by more than an order of magnitude.
}
 \end{figure}

In this appendix we first discuss a perturbative analytic expression for the deflection-induced contribution to the mean field and then the tricks we used to reduce the number of simulations needed to calculate the mean field.
We use the $\star$ operator for multiplication of infinite dimensional matrices across the survey area (continuous sky indices). In the isotropic limit this multiplication reduces to a standard convolution. Sums over discrete indices (CMB pixels, or Stoke fields) are indicated by juxtaposition. In particular, the weight matrices $W$ of the quadratic estimators act with field and pixel indices on the right, and with field and sky indices on the left. The beam operation $B$ maps the Stokes sky onto the data Stokes pixelization. Hence it acts on field and sky indices on the right, and field and pixel indices on the left.

\subsection{Deflection-induced mean-field contribution}
A good handle on the deflection-induced contribution to the mean field can be obtained perturbatively. 
We start from
\beq
g^{\rm MF}_a(\vecx) = \frac 12\frac{\delta \ln \det \Cov_\deflect}{\delta \normdeflect_a(\vecx)},
\enq
where unlike the previous appendix, we do not need to distinguish between upper and lower indices on the flat sky.

Recall that $\vecx$ refers to an arbitrary point on the sky, unrelated to the pixelization: we view each element of the covariance matrix as a functional of the deflection field. The linear response of the mean field is the second variation of the log-determinant functional. Hence,
\beq
\frac{\delta g^{\rm MF}_a(\vecx)}{\delta \normdeflect_b(\vecy)} =- \frac 12 \Tr \: \Cov^{-1}_\deflect \frac{\delta \Cov_\deflect}{\delta \normdeflect_a(\vecx)}\Cov^{-1}_\deflect \frac{\delta \Cov_\deflect}{\delta \normdeflect_b(\vecy)}   +\frac 12 \Tr \: \Cov^{-1}_\deflect \frac{\delta^2 \Cov_\deflect}{\delta \normdeflect_a(\vecx) \delta \normdeflect_b(\vecy)}.
\enq
The first term (identical to the likelihood Fisher matrix), when evaluated at zero displacement, is simply minus the inverse $N^0_L$ lensing quadratic estimator response (evaluated with unlensed weights). It is convenient to introduce $\xi$, the real-space two-point function of the unlensed CMB fields, and $\xi_{,a}$, its derivative with respect to coordinate axis $a$. For each element of the covariance matrix, we may then write
\beq
\left. \frac{\delta \Cov_\deflect}{\delta \normdeflect_a(\vecx)} \right|_{\deflect=0}= B(\vecx) (\xi_{,a}\star B^\dagger)(\vecx) - (B \star \xi_{,a})(\vecx)B^\dagger(\vecx).
\enq
In this equation a sum over Stokes-field indices is implicit, and the pixel and further field indices are omitted on both sides in order to prevent visual cluttering. The second variation becomes
\beq
\left. \frac{\delta^2 \Cov_\deflect}{\delta \normdeflect_a(\vecx)\delta \alpha_b(\vecy)} \right|_{\deflect=0}= \delta^D(\vecr) \lb B(\vecx) (\xi_{,ab}\star  B^\dagger)(\vecy) + (B \star \xi_{,ab})(\vecy) B^\dagger(\vecx) \rb
 -  B(\vecx) \xi_{,ab}(\vecr) B^\dagger(\vecy) - B(\vecy) \xi_{,ab}(\vecr) B^\dagger(\vecx),
\enq
where $\delta^D$ is the Dirac delta function.
Introducing the isotropic operator
\beq
K(\vecx-\vecy) \equiv \lb B^\dagger \Cov_{\deflect =0}^{-1} B\rb(\vecx - \vecy),
\enq
all explicit dependence on the pixelization has disappeared, and a \revision{short calculation gives}

\beq
\label{MFpred}
R_{ab}(\vecr)  \equiv \left.\frac{\delta g^{\rm MF}_a(\vecx)}{\delta \normdeflect_b(\vecy)}\right|_{\deflect = 0} = \revision{+} \Tr \lb \lp \xi_{,a}\star  K \rp(\vecr) \lp \xi_{,b}\star  K \rp(\vecr)+  K(\vecr) (\xi_{,a}\star K \star \xi_{,b})(\vecr)  -  K(\vecr)\xi_{,ab}(\vecr) +  \delta^D(\vecr)  ( K \star \xi_{,ab})(\vecr)  \rb.
\enq
In harmonic space, the last term is a constant, and ensures the response to the unobservable deflection monopole $R_{ab}(\vecell = 0)$ vanishes as it should. The inverse of minus the first two terms is the usual $N^0$ lensing bias (here displayed with unlensed weights, and before projection onto gradient and curl components). For low noise experiments, $\xi K \xi \sim \xi$ on most scales, causing large cancellations between the second and third terms.

By design, we have thus
\beq
g^{\rm MF}_a(\vL) = \sum_b R_{ab}(\vL) \normdeflect_b(\vL) + O(\deflect^2),
\enq
where all terms can easily be calculated with a series of Fourier transforms.


Fig.~\ref{fig:MFpred} shows the expected contribution of the deflection mean field for the polarization reconstruction considered in Sec.~\ref{Sec:maskedrecon}, but with no sky cuts so that the deflection is the only mean-field source. The blue line shows the mean-field spectrum estimate obtained by averaging 500 simulations.
As a test case we used an input deflection field $\nabla\phi_0$, with spectrum
\beq
C^{\phi_0 \phi_0}_L = \frac{\lp C^{\phi\phi}_L\rp^2}{C^{\phi\phi}_L + N^0_L},
\label{WienerPhiSpectrum}
\enq
equivalent, from Eq.~\ref{start}, to the spectrum of the reconstruction expected at the first iteration step. The orange curve shows the predicted spectrum from Eq.~\ref{MFpred}, which is in very good agreement given the MC noise of the mean-field estimation, visible at high multipoles.



\subsection{Mean field tricks}
To improve the calculation of the mean field, we introduce two tricks that together can reduce the MC noise power in the mean-field estimate by orders of magnitude.
The first, and often the most powerful, is to modify the weights used in the quadratic estimator that is averaged to calculate the mean field from the simulations. The usefulness of this trick, however, depends on the adequacy of $\Cov_\deflect$ to represent the true covariance of the data. Specifically, it makes use of
\beq
\label{covID}
\widetilde\Cov_\deflect^{-1} \equiv \Cov^{-1}_\deflect \av{\Stdat\Stdatdag}\Cov^{-1}_\deflect =\Cov^{-1}_\deflect .
\enq
In practice, $\Cov_\deflect$ input to the likelihood is always going to be  only an approximation to the true unknown data covariance. Accurate timestream simulations $\Stdat$ can sometimes be used to quantify the mean field (and other biases) for the standard quadratic estimator. But moderate numbers of simulations do not directly provide the means to accurately calculate $\Cov_\deflect$ or apply its inverse. The simulations themselves may of course also only capture only parts of the complexity entering the relevant systematics and the data processing. These are difficulties affecting the quadratic estimator as well, not specifically the iterative scheme proposed in this paper, that requires some amount of testing in a realistic situation.


From Eq. \eqref{MF}, the mean field satisfies
\beq
g^{\rm MF}_a(\n)  = \av{g_a^{\textrm{QD}}(\n)} = \av{[(V_{\deflect}\Stdat)]^i(\n) [W^a_{\deflect}\Stdat]_i(\n)}.
\enq
Performing the average using the normal weights of Eq.~\eqref{optweights} and assuming we can use Eq.~\ref{covID} results in
\beq
g^{\rm MF}_a(\n) = \Tr\: H^a_\deflect(\n,\n) \textrm{   with   } H^a_\deflect \equiv  B^\dagger \Cov_\deflect^{-1}B \star \D \:\nabla_a C^{\rm unl} \D^\dagger,
\enq
where the trace is over field indices.
Only the diagonal of $H^a_\deflect$ is relevant for the mean field, both in respect to sky and field indices,
and since $H^a$ is a vector it vanishes for isotropic fields where there is no mean field.

It is possible to construct mean-field estimators with modified weights, or acting on different maps: as long as the expectation of the quadratic estimator remains the same they will produce unbiased estimates of the mean field. We choose weights to apply to independent unit variance Gaussian variables $s$, with $\av{ss^\dagger} = {\rm diag}(1)$ on the  unmasked pixels, and use the following pair of weights
\beq
\label{Acc}
\begin{split}
W_1 = B^\dagger \quad \quad W^a_{2} =\D \nabla_a C^{\rm unl}_{\ell} \D^\dagger \star B^\dagger  \:\Cov_\deflect^{-1}.
\end{split}
\enq
The matrix inverse can be performed
in the same way as the usual filtering. We now proceed to justify this choice and explain why it has lower variance (though is not exactly minimum variance).

Ideally, we would like to choose a pair of weights $W_1, W^a_2$ to minimize the Gaussian MC noise on the mean field estimate while keeping the constraint $[W_1\av{s s^\dag} W^a_2{}^\dagger] (\vx,\vx) = \Tr H_\deflect^a(\vx,\vx)$.
The real-space MC noise covariance is
\beq
\begin{split}
N_0(\vecr) = (W_1  W_1^\dagger)(\vecr) (W_{2 }^a W_2^a{}^{\dagger})(\vecr) + (W_1 W_2^a{}^{\dagger})(\vecr) (W_{2}^a W_1^\dagger)(\vecr).
\label{MCNoiseVar}
\end{split}
\enq
For simplicity, we use the constraint
$W_1\av{s s^\dag} W^a_2{}^\dagger = W_1W^a_2{}^\dagger = H^a$,
which is much more stringent than only matching the diagonal but makes things more tractable. We then consider minimizing the variance in the isotropic limit,
in which case $W_1W^a_2{}^\dagger$ becomes a real-space convolution and
hence the constraint means that in harmonic space $W^a_2(\vecell) = H^a_\deflect(\vecell) / W_1(\vecell)$, where $W_1(\vecell)$ is a free function of multipole.
 A natural measure to minimize is the integrated variance from the above equation
\beq
\frac 1 V \sum_\vecL N^0_\vecL = N_0(\vecr = 0).
\enq
Using the constraint equation, minimizing $N_0(\vecr =0)$ gives
\beq
W_1(\vecell) \propto W_2^a(\vecell) H^a_\deflect(\vecr =0) \,,\qquad W_2^a(\vecell) \propto W_1(\vecell) H^a_\deflect(\vecr =0),
\enq
so the two weight functions are proportional. From the constraint this implies that $W_1(\vecell) \propto \sqrt{H^a_\deflect(\vecell) H^a_\deflect(\vecr =0)}$.

The key message of this calculation is that to get low MC noise the scale dependence of the weight functions should be similar,
though once isotropy is broken by the deflection in practice it is never possible to obtain exact square roots even in the absence of non-ideal effects.
%
Other aspects matter as well, such as how well the unavoidable matrix inversion behaves with the chosen weights, and the contribution of the non-ideal effects to the final MC-noise floor. The form given in Eq.~\ref{Acc} is an empirical compromise between these considerations, that we found works well in practice. It equilibrates the weights in a very crude way, simply by having the same powers of the signal on each leg (though not the exact scale dependence). Despite being crude, it reduces the MC noise floor by more than one order of magnitude for the polarization reconstruction performed in the main text. This is shown on Fig.~\ref{fig:MFpred}. The dash-dotted black line the MC noise of the estimator given by Eq. \ref{Acc}, and should be compared to the naive estimate MC noise given by the $N_0$ curve  (black, solid). Both curves were calculated with the lensed spectra weights, which is a slightly conservative estimate choice as the iterative search converges towards the optimal solution.
We found empirically that further modifying the weights to make their scale-dependence closer (for example multiplying $W_1$ by $\sqrt\ell$ and $W_2$ by $1/\sqrt\ell$ to equilibrate power in the no-beam, no-lensing, no-noise limit) did not give substantial further improvements.

The second trick is to subtract from each MC estimate the same estimate but based on an isotropic approximation to the posterior that is a close as possible to the true one, as described in Sec.~\ref{Sec:MF}.
\clearpage

\newpage
\twocolumngrid

\bibliography{julien,antony,cosmomc,lensingbib}

\end{document}

%% file: macros.tex
\newcommand{\la}{\langle}
\newcommand{\ra}{\rangle}

\newcommand{\SO}{\textit{Simons Observatory}}

\newcommand{\GNILC}{GNILC}
\newcommand{\herschel}[0]{\textit{Herschel}}
\newcommand{\mksym}[1]{\ifmmode {\rm #1}\else #1\fi}

\providecommand{\Planck}{\textit{Planck}}
\providecommand{\planck}{\Planck}

\providecommand{\text}[1]{\rm{#1}}

\providecommand{\arcmin}{{\rm arcmin}}

\providecommand{\CAMB}{{\tt camb}}

\providecommand{\HEALpix}{{\tt HEALpix}}

\newcommand{\begm}{\begin{pmatrix}}
\newcommand{\enm}{\end{pmatrix}}

\newcommand\ba{\begin{eqnarray}}
\newcommand\ea{\end{eqnarray}}
\newcommand\bea{\begin{eqnarray}}
\newcommand\eea{\end{eqnarray}}

\newcommand\be{\begin{equation}}
\newcommand\ee{\end{equation}}

\providecommand{\Tr}{\text{Tr}}

\newcommand{\boldvec}[1]{{\mbox{\boldmath{$#1$}}}}

\newcommand{\vL}{\boldvec{L}}

\newcommand{\ve}{\boldvec{e}}

\newcommand{\vn}{\boldvec{n}}

\newcommand{\vx}{\boldvec{x}}